\documentclass[journal]{IEEEtran}
\pdfoutput=1

\usepackage{booktabs} 
\usepackage{color}
\usepackage[dvipsnames]{xcolor}

\usepackage{subfigure}
\usepackage{amsmath,amsfonts,color,graphics,graphicx}
\usepackage[T1]{fontenc}
\usepackage{algorithm,algorithmic,url}
\usepackage{multirow}

\newcommand{\D}{{\ensuremath{\mathcal D}}}
\newcommand{\A}{{\ensuremath{\mathcal A}}}
\newcommand{\M}{{\ensuremath{\mathcal M}}}
\newcommand{\X}{{\ensuremath{\mathcal X}}}

\newcommand{\GT}{{\text{GT}}}
\newcommand{\EGT}{{\text{EGT}}}
\newcommand{\TT}{{\text{TT}}}
\newcommand{\LBT}{{\text{LBT}}}
\newcommand{\EE}{{\text{EE}}}
\newcommand{\RGI}{{\text{RGI}}}
\newcommand{\AGI}{{\text{AGI}}}

\newcommand{\ignore}[1]{}

\newtheorem{definition}{Definition}

\DeclareMathOperator*{\argmin}{arg\,min} \DeclareMathOperator*{\argmax}{arg\,max} 

\begin{document}

\title{Metrics Towards Measuring Cyber Agility}

\author{
Jose David Mireles, Eric Ficke, Jin-Hee Cho,~\IEEEmembership{Senior Member,~IEEE}, Patrick Hurley, and Shouhuai Xu
\thanks{J. Mireles, E. Ficke, and S. Xu are with the Department of Computer Science, University of Texas at San Antonio. 
J.H. Cho is with the Department of Computer Science, Virginia Tech.
P. Hurley is with the U.S. Air Force Research Laboratory, Rome, NY.
Correspondence: {\tt shxu@cs.utsa.edu}}
}

\maketitle

\begin{abstract}
In cyberspace, evolutionary strategies are commonly used by both attackers and defenders. For example, an attacker's strategy often changes over the course of time, as new vulnerabilities are discovered and/or mitigated. Similarly, a defender's strategy changes over time. These changes may or may not be in direct response to a change in the opponent's strategy. In any case, it is important to have a set of quantitative metrics to characterize and understand the effectiveness of attackers' and defenders' evolutionary strategies, which reflect their {\em cyber agility}. Despite its clear importance, few systematic metrics have been developed to quantify the cyber agility of attackers and defenders. In this paper, we propose the first metric framework for measuring cyber agility in terms of the effectiveness of the dynamic evolution of cyber attacks and defenses. The proposed framework is generic and applicable to transform any relevant, quantitative, and/or conventional static security metrics (e.g., false positives and false negatives) into dynamic metrics to capture dynamics of system behaviors. In order to validate the usefulness of the proposed framework, we conduct case studies on measuring the evolution of cyber attacks and defenses using two real-world datasets. We discuss the limitations of the current work and identify future research directions.
\end{abstract}


\IEEEpeerreviewmaketitle

\section{Introduction} \label{sec:introduction}
\IEEEPARstart{I}{n} order to maximize the effectiveness of cyber attacks or defenses, both cyber attackers and defenders frequently evolve their strategies. The rule-of-thumb is that cyber attackers are more agile in adapting their strategies than cyber defenders, because cyber defenders often tend to take \emph{reactive} responses to new attacks. Accordingly, cyber attack incidents are frequently reported in news media. However, the state-of-the-art technology does not provide quantitative metrics that can measure how well cyber attackers or defenders are able to adapt or update their resources over time. We call this problem \emph{measuring cyber agility}. Cyber agility and its quantification have recently been  recognized as critical cybersecurity issues that are little understood~\cite{McDaniel:2014:SSA:2663474.2663476,marvel2015framework,Cho-Milcom-2016,Pendleton16,XuSTRAM2018}.

In this paper, we take a first step towards tackling the problem of measuring cyber agility. We propose a systematic set of quantitative metrics to measure cyber attack and defense \emph{evolution generations} (or simply \emph{generations}), which are cyber attack and defense updates that can be considered as ``building-blocks'' or ``atomic moves'' used by cyber attackers and defenders in their operational practice. The notion of generations is important because cyber attacks and defenses often evolve over time. It is also important to see that the causality of evolution generations may differ between them. For example, some evolution generations are caused by specific opponent moves, but others aren't (e.g., moving-target defense is not necessarily caused by any specific attacks). 
Intuitively, we can quantify the relative agility of cyber attackers and defenders by characterizing their evolving strategies and measuring their effectiveness during the course of interplay between cyber attackers and cyber defenders. 

The aforementioned \emph{dynamic} view of cybersecurity metrics, which we pursue in the present paper, contrasts with the conventional \emph{static} view of cybersecurity metrics as follows: the dynamic view reflects a system's evolution over a period of time, while the static view captures measurements of metrics at a certain time point or in an aggregated way. To the best of our knowledge, this is the first work that defines systematic metrics to quantify the effectiveness of cyber attack and defense evolution gearing towards measuring cyber agility.

\subsection{Key Contributions} \label{subsec:contributions}
This work makes the following key contributions:
\begin{itemize}
	\item {\bf Development of a system-level evolutionary metric framework}: We develop a dynamic metric framework that can deliver more information of system behaviors than static metrics. There exist some \emph{time-dependent} metrics that measure the outcome of attack-defense interactions at \emph{every time point} $t$ \cite{XuTAAS2012,XuIEEETDSC2012-multivirus,XuInternetMath2012,XuGlobalStability2016,XuTAAS2014,XuInternetMath2015ACD}.  However, our proposed metric framework takes one step beyond those time-dependent metrics because it is capable of measuring, at time $t$, the system state at both a past time $t'$ (where $t'<t$) and a future time $t''$ (where $t''>t$). This means that the framework can be used for \emph{retrospective security analysis} (e.g., identifying what a defender did right or wrong in the past). 
	In contrast, the time-dependent metrics \cite{XuTAAS2012,XuIEEETDSC2012-multivirus,XuInternetMath2012,XuGlobalStability2016,XuTAAS2014,XuInternetMath2015ACD} only measure the defense effectiveness at $t$ when the measurement is made at $t$. 
	\item {\bf Transformation of static metrics into dynamic metrics}: We transform conventional static metrics into dynamic metrics to measure attack and defense effectiveness at each evolution generation. For example, static metrics (e.g., false-positive or false-negative rate) can be transformed into dynamic metrics to capture a dynamic sequence of evolution generations by attackers or defenders. Thus, the framework can be viewed as a ``compiler'' that transforms static metrics into dynamic metrics.
	\item {\bf Validation of the proposed metric framework using real datasets}:
	In order to validate the framework, we apply it to analyze two real-world datasets, one is the network traffic collected at a honeypot instrument \cite{ProvosUsenixSecurity04} and the other corresponds to the DEFCON Capture The Flag (CTF) exercise~\cite{DEFCONDatasets}. We use the Snort intrusion detection system (IDS) \cite{RoeschLISA99} as a defense mechanism, whose detection capability are frequently updated (reflecting defense generations that may or may not be caused by specific attack evolution generations).
\end{itemize}
In the present study we focus on proposing a systematic framework with clear definitions of cyber agility metrics. Although our case study is limited by the datasets we have access to, we hope researchers having access to semantically richer datasets can applying our framework to their datasets to draw deeper insights towards taming cyber agility.


The remainder of the paper is organized as follows. Section~\ref{sec:related-work} discusses the background and related prior studies. Section~\ref{sec:framework} presents the proposed metric framework. Section~\ref{sec:discussion} discusses the insights obtained from Section \ref{sec:framework}. Section~\ref{sec:case-study} presents the case study on using the framework to analyze two real datasets. Section~\ref{sec:limitation} discusses limitations of this study and future research directions. Section~\ref{sec:conclusion} concludes this paper.

\section{Background \& Related Work} \label{sec:related-work}

\subsection{Concept of Agility and Agility Metrics}

Agility has been recognized as one of the key system metrics, but has been studied only in an ad-hoc manner \cite{McDaniel:2014:SSA:2663474.2663476, AlbertsAQ2014}. It has been investigated in multiple domains
\cite{Kidd94, Yusuf99, Sherehiy07, Alberts07, Conboy09, AlbertsCCRP2011, Salo16}. 
In an enterprise system, agility is defined as the ability to deal with sudden changes in an environment (e.g., the latency of response to sudden or unexpected changes~\cite{McDaniel:2014:SSA:2663474.2663476, Kidd94, Yusuf99, Sherehiy07, Salo16}). In the systems engineering domain, agility measures a system's capability of the reactive or proactive response to sudden environmental changes~\cite{Dove05}. 
In military settings, agility refers to the ability an entity takes an effective action under dynamic, unexpected environments that may threaten the system's sustainability~\cite{Alberts07}. For example, the \emph{qualitative} notion of \emph{agility quotient} is proposed to accommodate six attributes, including responsiveness, versatility, flexibility, resilience, adaptiveness, and innovativeness ~\cite{AlbertsCCRP2011}. 

In the cybersecurity domain, agility often refers to reasoned changes to a system or environment in response to functional or security needs, but has not been paid due attention until very recently
\cite{AlbertsAQ2014,McDaniel:2014:SSA:2663474.2663476,marvel2015framework,Cho-Milcom-2016,Pendleton16,XuSTRAM2018}.
While it would be intuitive to understand agility as how \emph{fast} (e.g., response ability ~\cite{Dove05}) and how \emph{effective} a system can adapt its configuration to unexpected attacks against it (e.g., considering the cost incurred by the degraded system performance or the cost incurred by response actions ~\cite{McDaniel:2014:SSA:2663474.2663476, DoveBook2001}), the concept of agility is, like other metrics, elusive to formalize. Indeed, there is no rigorous or quantitative definition of agility in the cybersecurity domain and existing attempts to model agility have not been able to produce concrete measurements \cite{marvel2015framework,hult2014good,morrisking2015analyzing}. Despite the apparent importance, systematic and \emph{quantitative} agility metrics have not been studied and understood in-depth. 

In this paper, we tackle the problem of measuring agility in the cybersecurity domain and propose an agility metric framework based on evolution generations of cyber attacks and defenses. To the best of our knowledge, this is the first systematic metric framework for understanding and measuring cyber agility.
Our framework is both general and flexible because it can accommodate attack and defense evolution generations that may or may not be incurred by specific opponent evolution generations.

\subsection{Dynamic Security Metrics vs. Agility Metrics}
The importance of security metrics and the challenges of developing useful security metrics have been recognized by security communities~\cite{IRC-hardproblemlist, NITRD,NSAHardProblemList, NIST800-55Rev1,CIS}. Most metrics proposed in the literature are \emph{static} in nature because they are often defined without considering dynamics over time~\cite{Pendleton16,XuSTRAM2018,DBLP:journals/tifs/DuSCCX18}. This implies that static metrics can capture either a system's snapshot at a particular time or a system's overall behavior/state for a period of time; this static view can easily overlook the evolution of attacks and/or defenses over the time horizon. For example, the effectiveness of anti-malware tools (e.g., the false positive rate or false negative rate) is often measured based on malware samples collected during a period of time (e.g., one year), while ignoring their instantaneous evolution over time. Taking one step further from  static metrics, \emph{time-dependent} security metrics have been studied to characterize and quantify system states at different times, such as the proportion of compromised computers in a network~\cite{XuTAAS2012,XuIEEETDSC2012-multivirus, XuInternetMath2012, XuGlobalStability2016,XuTAAS2014,XuInternetMath2015ACD,XuIEEEACMToN2019,XuInternetMath2015Dependence, XuHotSOS14-MTD, XuHotSoS15, XuGameSec13,XuQuantitativeSecurityHotSoS2014,XuCybersecurityDynamicsHotSoS2014,XuBookChapterCD2018,DBLP:conf/hotsos/ChenCX18,DBLP:conf/hotsos/ChenCX18a}.

The dynamic metrics proposed in this paper aim to capture a system's state, covering both a previous time $t'$ and a future time $t''$, when a measurement is made at $t$ where $t' < t < t''$. On the other hand, the time-dependent metrics measure the system's current state at time $t$, as mentioned in Section \ref{subsec:contributions}. For example, the effectiveness of an IDS can be characterized by its true-positive rate and/or false-negative rate, which may not stay the same over time because its decision engine (e.g., the rule set in the case of the Snort) is frequently updated.

There have been other proposals for dynamic security metrics ~\cite{Pendleton16}, including (i) metrics for measuring the strength of preventive defense (e.g., \emph{reaction time} between the observation of an adversarial entity at time $t$ and the blacklisting of the adversarial entity at time $t'$~\cite{DBLP:conf/raid/KuhrerRH14}); (ii) metrics for measuring the strength of reactive defense (e.g., \emph{detection time} between a compromised computer starting to wage attacks at times $t$ and $t'$ at which the attack is first observed by some cyber defense instrument~\cite{Rajab:2005:EDW:1251398.1251413,XuVineCopula2017}); (iii) metrics for measuring the strength of overall defense (e.g., \emph{penetration resistance} for measuring the level of effort that is imposed on a red team in order to penetrate into a cyber system~\cite{DBLP:conf/discex/Levin03,CybenkoIEEEComputer2008}); and (iv) metrics for measuring and forecasting cyber threats and incidents \cite{XuTIFS2013,XuTIFS2015,XuPLoSOne2015,XuJAS2016,XuJournalAppliedStatistics2018,XuTIFSDataBreach2018}. Although these metrics are related to time, they are geared towards individual security events. In contrast, our framework is systematic and correlates attack-defense over the horizon of time.

\section{The Metrics Framework} \label{sec:framework}
The proposed framework aims to define metrics to measure the effectiveness of attack and defense generations during the course of attack-defense interactions.

\subsection{Guiding Principles}
The framework is designed under the following principles:
\begin{itemize}
	\item \textbf{Leveraging static security metrics}: Although most existing security metrics measure the \emph{static} aspects of a system's security, some of them are well defined and commonly accepted, such as detection errors (e.g., false-positives or false-negatives) for an IDS or anti-malware system. The framework aims to accommodate static metrics that were defined in the past or may be defined in the future.
	
	\item \textbf{Considering the evolution of both attack and defense behaviors}: The framework aims to understand and improve defenders' evolution over the course of time. To this end, we will answer the following research questions:
	\begin{itemize}
		\item To what extent is an attacker or defender evolving based on its opponent's new strategy?
		\item Which evolutionary strategy is more effective in terms of an attacker's or defender's perspective?
		\item Which party (i.e., an attacker or defender) is more active over the course of attack-defense interactions?
	\end{itemize}
	\item \textbf{Identifying the core metrics to measure systems security}: Since the effectiveness of attack and defense generations may not be adequately reflected by a single metric, we consider a suite of metrics that measure evolution generations from multiple perspectives. These metrics may be then aggregated using an appropriate method (e.g., a weighted average).
	\item \textbf{Coping with new or zero-day attacks}: Current defenses have a very limited power in detecting new or zero-day attacks. For a clear understanding on the effectiveness of attack and defense evolutions, we measure generations by tracing recorded network traffic and/or computer execution. This allows us to characterize defense failures in retrospect.
\end{itemize}

\subsection{Representation of Attacks and Defenses over Time}

In this paper, the term ``target system'' is used to represent a range of systems, from a single computer (or device) to an enterprise network or a cloud. A target system is defended by human defenders who can use a variety of defense tools. 
Therefore, ``attackers" and ``defenders" refer to either humans,  automated systems, or a combination of them (depending on the practice). The term ``attack and defense generations'' is used to refer to the atomic evolution of behaviors by attackers and defenders. 

Specifically, we consider time horizon $t\in [0,T]$, where $T$ can be infinite in theory (i.e., $T=\infty$) but is often finite in practice (i.e., $T<\infty$). Since most metrics are often measured at discrete times (e.g., daily or hourly), we consider discrete-time over $t=0,1,\ldots,T$. That is, we treat each generation as if it happens at an instant time in the beginning of each time unit (e.g., day or hour). In practice, it is possible that attacks and defenses are respectively observed over time intervals, say, $[t_1,t_2]$ and $[t'_1,t'_2]$, where $t_1\neq t'_1$ and $t_2\neq t'_2$. In this case, we can treat $\min(t_1,t'_1)$ as time 0 and $\max(t_2,t'_2)$ as time $T$.

We use the term ``defense" at discrete time $t$ to refer to the defense tool and the human defender(s) employed at time $t$. This is important because defense tools may be updated with newer versions and human defenders may join/leave a defense team at any point in time. {Such a change in defense produces a new defense ``generation''.} We denote the defense at time $t$ by $\D_t$ , where $\D_0,\D_1,\ldots,\D_T$  represent the evolution of the defense over time $t=0,1,\ldots,T$. Note that $\D_t=\D_{t+1}$ indicates that the defense at time $t$ and the defense at time $t+1$ are the same and therefore belong to the same defense generation. Note also that $\D_t$ can be described by a nominal scale, such as a version number for a defense tool or an action by a human defender. In this way, we can check whether or not the defense at time $t$ {belongs to a different generation than} the defense at time $t+1$.
For example, suppose a target system's defense is started at $t=0$ (e.g., Jan. 1). Suppose the time unit is a day and the defense at $t=0$ consists of a human defender and an attack-detection tool with version 10.1.2. Suppose the attack-detection tool is updated on the first day of each month. This means that $\D_0=\D_1=\ldots=\D_{30}\neq \D_{31}$, where $\D_{31}$ refers to the same human defender but the attack-detection tool is version (say) 10.1.3 on Feb. 1.

Similarly, we use the term ``attack'' to describe attack tools, attack tactics, or attack vectors as well as the human attackers exploiting these attack tools, attack tactics, or attack vectors. Let $\A_{t}$ denote an attack against a target system at time $t$, and $\A_0,\A_1,\ldots,\A_T$ represent the evolution of the attack over $t=0,1,\ldots,T$. Note also that $\A_t$ can be measured by a nominal scale. Hence, we can detect if two attacks performed at two different points in time belong to the same generation.

{\bf Remark.} In the discrete-time model, attack and defense generations are assumed to evolve at discrete, \emph{deterministic} time points. In practice, generations can evolve over \emph{stochastic} time, implying that the time resolution should be sufficiently small. When the monitoring time interval is infinitely small, a continuous-time model should be used. In reality, the highest time resolution is what can be measured by a computer clock. The investigation on whether to consider a continuous-time model or not is beyond the scope of this paper.

\subsection{Representation of Defense Effectiveness over Time}

The effectiveness of defense $\D_t$ against attack $\A_t$ at time $t\in [0,T]$ is measured by some \emph{static} metrics (e.g., false-positive rate or false-negative rate). A metric $M$ is a mathematical function that maps the target system (or a particular property or attribute of the target system) at time $t$ to a value in a range (e.g., false-positive rate)~\cite{Pendleton16}. 
In order to make the presentation succinct, we assume that the range of any metric $M \in \M$ can be normalized to $[0,1]$, where a larger value is more desirable from a defender's point of view (e.g., a larger value means a higher level of security). Some metrics with an opposite meaning (e.g., a smaller false-positive rate, denoted by $fp$, is better) can be adjusted to be consistent with the scale in $M$ (e.g., using $1-fp$ instead of $fp$).

Let $\M$ denote the universe of static metrics, including both existing metrics and metrics that are yet to be defined. For a metric $M \in \M$, we use $\D_t(\A_t,M)$ to denote the effectiveness of defense $\D_t$ against attack $\A_t$ at time $t\in [0,T]$ in terms of metric $M$. 
By considering metric $M \in \M$ over time $t=0,1,\ldots,T$, we obtain a sequence of effectiveness measurements, which can be leveraged to define cyber defense agility as described below. 

Table \ref{table:notations} summarizes the main notations used in this paper.

\begin{table}[!htbp]
	\centering
	\caption{Summary of {key} notations and their meanings.\label{table:notations}}
	\vspace{-2mm}
	\begin{tabular}{|l|p{.35\textwidth}|}
		\hline
		Notation& Description \\ \hline
		$[0,T]$ & Time horizon, $t \in \{0,1,\ldots,T\}$ \\ \hline
		$\D_t$ & Defense at time $t\in [0,T]$ \\ \hline
		$\A_{t'}$ & Attack at time $t'\in [0,T]$ \\ \hline
		$\X$ & Attacker or defender\\ \hline
		$\ell$ ($\X$)  & Number of generations made by $\X$ during $[0,T]$ \\ \hline
		$\M$ & Universe of static security metrics for measuring defense effectiveness, scaled in $[0,1]$ \\ \hline
		$\D_t(\A_{t'},M)$ & Effectiveness of defense $\D_t$ at time $t$ against attack $\A_{t'}$ at time $t'$ in terms of metric $M\in \M$\\ \hline
		$\GT(\D,t)$     & Defender's \emph{Generation-Time} at time $t$\\ \hline
		$\GT(\A,t')$     & Attacker's \emph{Generation-Time} at time $t'$\\ \hline
		$\EGT(\D,t)$    & Defender's \emph{Effective-Generation-Time} at time $t$\\ \hline
		$\EGT(\A,t')$    & Attacker's \emph{Effective-Generation-Time} at time $t'$\\ \hline
		$\TT(\D,t)$     & Defender's \emph{Triggering-Time} at time $t$ \\ \hline
		$\TT(\A,t')$     & Attacker's \emph{Triggering-Time} at time $t'$ \\ \hline
		$\LBT(\X)$    & \emph{Lagging-Behind-Time} of $\X$ \\\hline
		$\EE(\D,t)$    & Defender's \emph{Evolutionary-Effectiveness} at time $t$\\\hline
		$\EE(\A,t')$    & Attacker's \emph{Evolutionary-Effectiveness} at time $t'$\\ \hline
		$\RGI(\X)$    & \emph{Relative-Generational-Impact} of $\X$ \\\hline
		$\AGI(\X)$    & \emph{Aggregated-Generational-Impact} of $\X$ \\\hline
	\end{tabular}
\end{table}

\subsection{Example Scenario}

\begin{figure}[!htbp]
	\centering
	\subfigure[Attack generations vs. defense generations: A solid-unfilled circle indicates the start of a defense generation, a solid-filled circle indicates the start of an attack generation, and a dashed-unfilled circle indicates no change by the defender or attacker.]{\includegraphics[width=.4\textwidth]{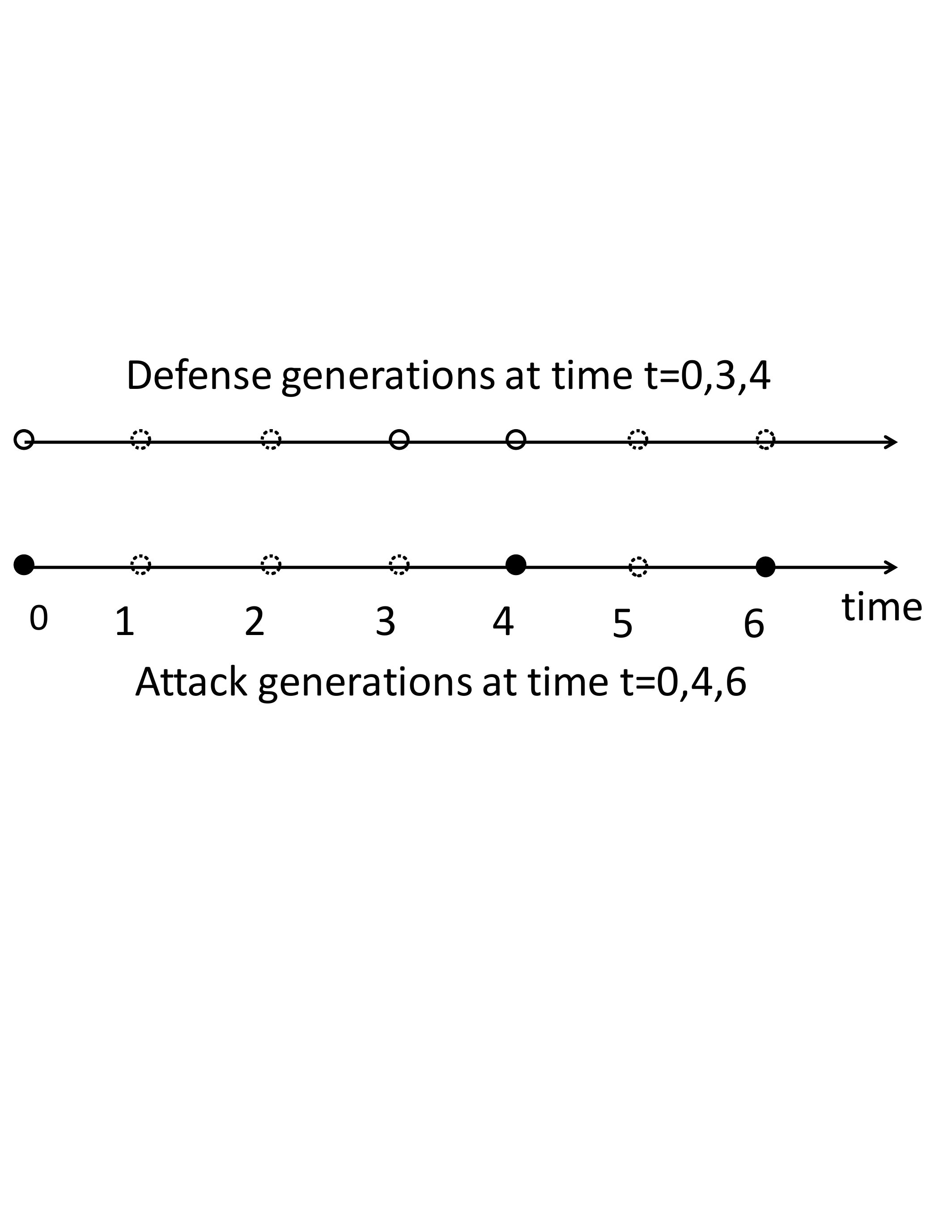} \label{fig:framework-a}}
	\vspace{-3mm}
	\subfigure[Attack-defense generations: `$\otimes$' indicates that both defense and attack evolve to new generations, a solid-unfilled circle indicates a change in defense generation but not in attack generation,
	a solid-filled circle indicates a change in the attack but no change in the defense, and a dashed-unfilled circle indicates no change in both defense and attack.]{\includegraphics[width=.4\textwidth]{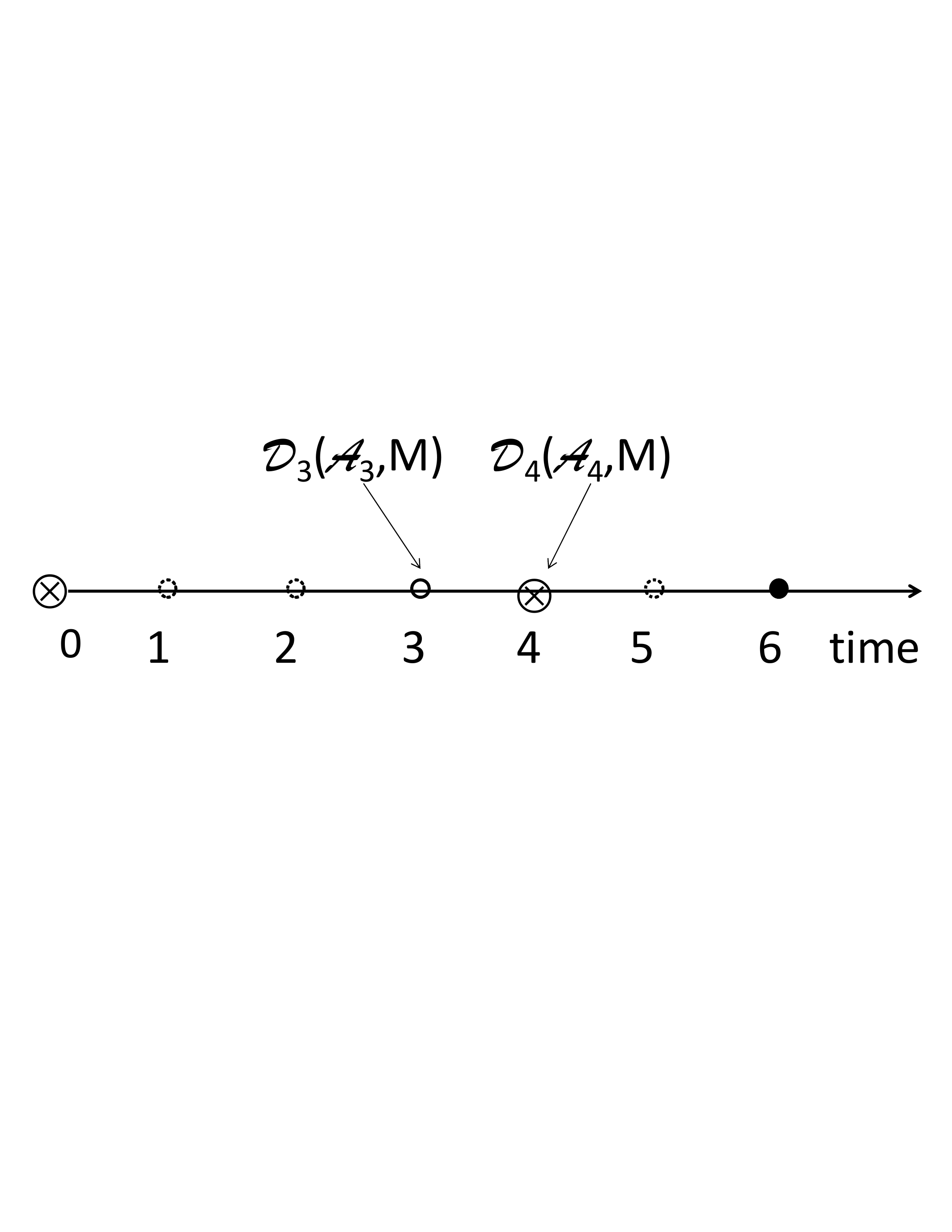}\label{fig:framework-b}}
	\hfill
	\caption{An example of attack and defense generations over $t \in [0, 6]$.}
	\label{fig:framework}
\end{figure}

Fig. \ref{fig:framework} shows an example used throughout the rest of this section. Fig. \ref{fig:framework} (a) illustrates that a defender evolved (e.g., updated a security software version) at $t=0, 3$ {and $4$} while making no changes at $t=1, 2, 5$ {and $6$}. An attacker evolved (e.g., changed an attack strategy) at $t=0, 4$ {and $6$} while making no changes at $t=1, 2, 3$ {and $5$}.
Fig. \ref{fig:framework} (b) views attack-defense generations at the same time scale axis, by  demonstrating how the attacker's evolution times may or may not coincide with the defender's evolution times.

\subsection{Overview of Metrics}
\begin{figure*}[htbp!]
	\centering
	\includegraphics[width=.76\textwidth]{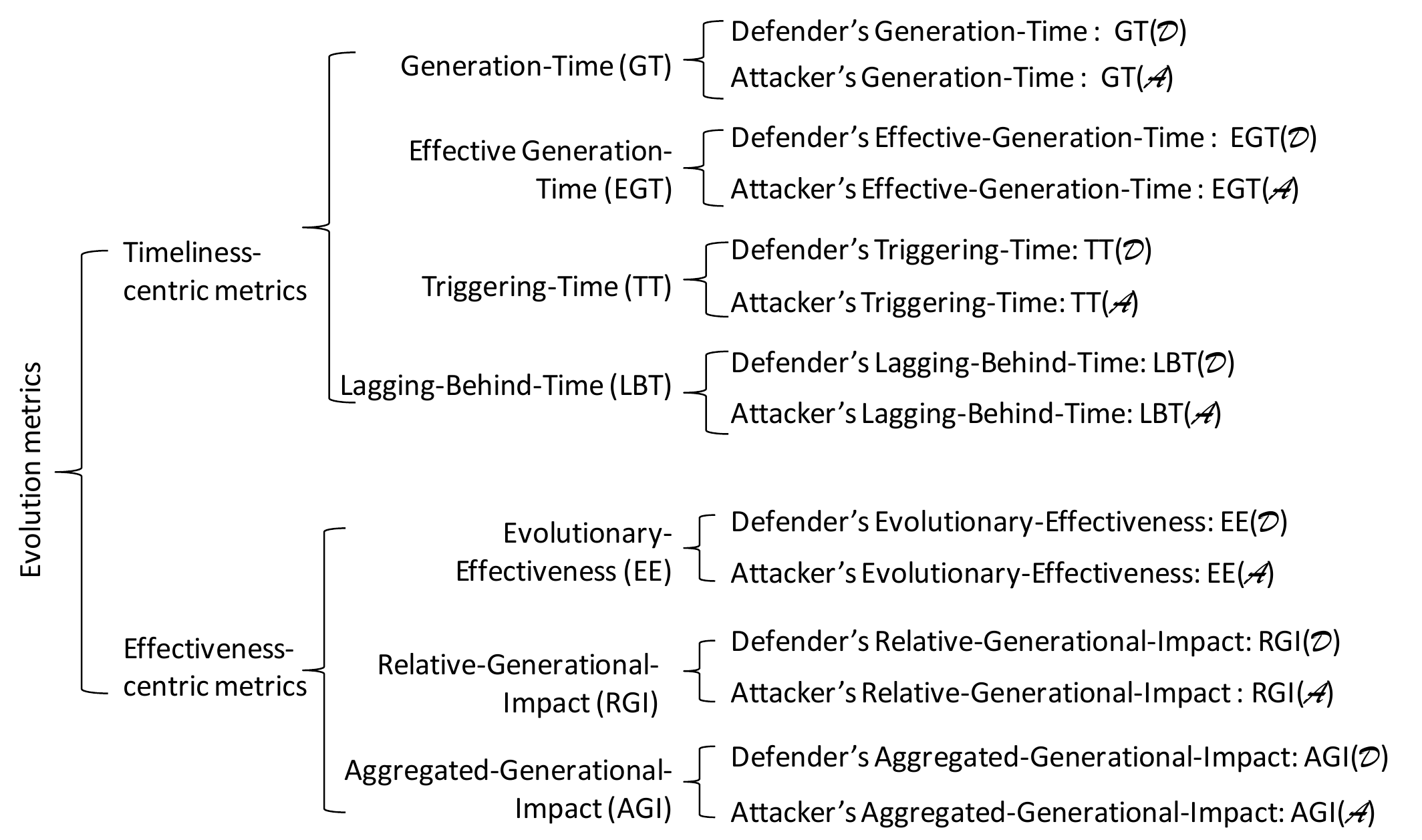}
	\caption{Overview of metrics with respect to attackers and defenders, where $\RGI(\D)=-\RGI(\A)$ and $\AGI(\D)=-\AGI(\A)$ but this kind of relationship does not apply to the other metrics. }
	\label{fig:framework-final}
\end{figure*}

At a high level, we consider two dimensions of evolution: \emph{timeliness} and \emph{effectiveness}. Timeliness reflects the time it takes to evolve new generations while effectiveness reflects impacts of these generations.
However, \emph{timeliness-oriented} metrics can use effectiveness as a reference, and \emph{effectiveness-oriented} metrics can use time as a reference.
Fig. \ref{fig:framework-final} summarizes these metrics and the structural relationship between them.

\subsubsection{\bf Timeliness-Oriented Metrics}
This suite of metrics measures how \emph{quickly} one adversarial party ({i.e., an} attacker or defender) evolves its strategies with or without considering the resulting effectiveness. This suite contains 4 metrics, which are equally applicable to both an attacker and defender, leading to 8 metrics in total. The 4 metrics are as follows:
\begin{itemize}
	\item \emph{Generation-Time} (GT) measures the time between two consecutive generations of strategies that are observed by the measuring party (i.e., an attacker or defender).
	\item \emph{Effective-Generation-Time} (EGT) measures the time it takes for a party to evolve a generation which indeed increases the effectiveness against the opponent.
	\item \emph{Triggering-Time} (TT) measures the length of time since the opponent's reference generation that (if observed) may have triggered a particular generation.
	\item \emph{Lagging-Behind-Time} (LBT) measures how long a party lags behind its opponent with respect to a reference time.
\end{itemize}
These 4 metrics are \emph{random variables} in nature, sampled over the time horizon $[0,T]$.

\subsubsection{\bf Effectiveness-Oriented Metrics}
This suite of metrics {measures} the effectiveness of generations over the course of the evolution. This suite contains 3 metrics, which are equally applicable to both an attacker and defender, leading to 6 metrics in total. The 3 metrics are as follows:
\begin{itemize}
	\item \emph{Evolutionary-Effectiveness} (EE) measures the overall effectiveness of generations with respect to the opponent's generation. This is a \emph{random variable} over $t\in [0,T]$.
	\item \emph{Relative-Generational-Impact} (RGI) measures the effectiveness gained by generation $i$ over {that} of generation $i-1$.
	\item \emph{Aggregated-Generational-Impact} (AGI) measures the gain in the effectiveness of all generations $t\in [0,T]$.
\end{itemize}

\subsubsection{\bf Relationship between the Metrics}

\begin{figure}[htbp!]
	\centering
	\includegraphics[width=.49\textwidth]{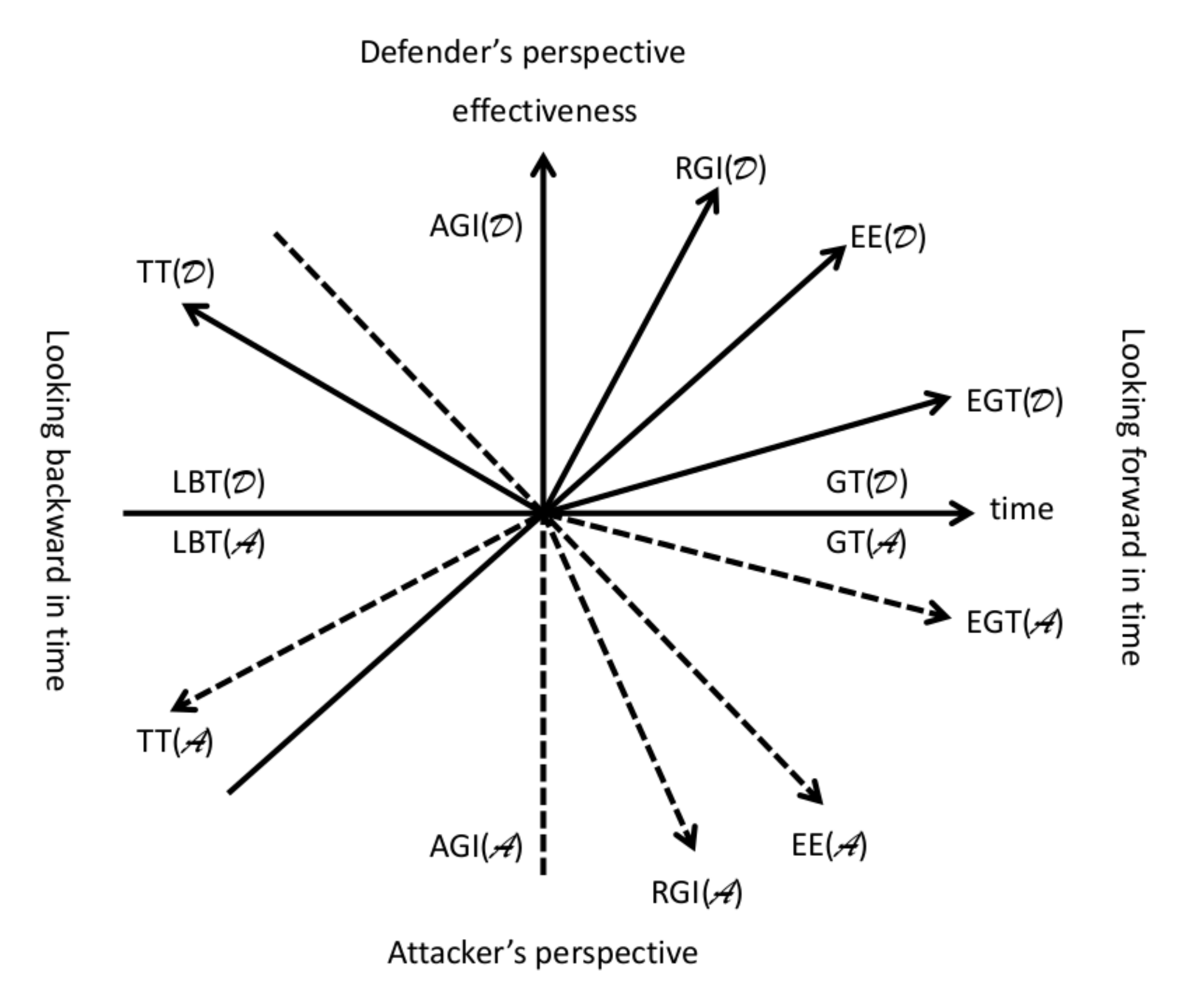}
	\caption{The structural relationship between the aforementioned 14 metrics according to the time and effectiveness dimensions. Intuitively, some metrics focus on time ($x$-axis) and look either forward or backward in time, while others are more oriented towards effectiveness ($y$-axis), where the defender's metrics are reflected across the $x$-axis to create the attacker's perspective in dashed lines. 
		Note that $\EE(\D)$ is defined in Quadrants 1 and 3 {and} $\EE(\A)$ is defined in Quadrants 2 and 4, {while} the other metrics are defined in a single Quadrant.}
	\label{fig:metrics-systematization}
\end{figure}

Fig. \ref{fig:metrics-systematization} systematizes the relationship between the aforementioned 14 metrics. These metrics are organized in two dimensions: \emph{time} ($x$-axis) and \emph{effectiveness} ($y$-axis). The metrics in the upper half of the plane (i.e., $y>0$) represent the defender's perspective; the metrics in the lower half of the plane (i.e., $y<0$) represent the attacker's perspective. The metrics in the right-hand half of the plane (i.e., $x>0$) look forward in time based on some reference point; the metrics in the left-hand half of the plane (i.e., $x<0$) look backward in time based on some reference point.

A metric closer to the $x$-axis indicates a more time-oriented perspective, while the other metrics {are} more oriented towards an effectiveness perspective. A metric on the $x$-axis is defined primarily based on the time dimension, including $\GT$ and $\LBT$. Effectiveness-oriented metrics are $\RGI$ and $\AGI$, where $\AGI(\D)=-\AGI(\A)$. The metrics defined from the defender's perspective and the metrics defined from the attacker's perspective are symmetric across the $x$-axis. $\EE(\D)$ and $\EE(\A)$ are defined over the entire plane because they look both backward and forward in time.

\subsection{Timeliness-Oriented Metrics}

\subsubsection{\bf Generation-Time (GT)} This metric measures the time it takes for a party to evolve its strategy, which may or may not be induced by the opponent's evolution in strategy. 
GT is a \emph{random variable} because generations often evolve based on some stochastic events.

{\bf Defenders' GT, denoted by $\GT(\D)$:}  Suppose the defense is evolved at $t_0=0,t_1,\ldots,t_\ell \leq T$, namely $\{t_0,t_1,\ldots,t_\ell\}\subseteq [0,T]$. The defender's GT, namely random variable $\GT(\D)$, is sampled by $\GT(\D,i)$'s, where
\begin{equation}
\GT(\D,i)=t_{i+1}-t_i ~~\text{for}~~i=0,1,\ldots,\ell-1.
\end{equation}
This implies that $\D_{t_i+\Delta t} = \D_{t_i}$ for any $\Delta t < t_{i+1}-t_i$ and $\D_{t_i+\GT(\D,i)} = \D_{t_{i+1}} \neq \D_{t_i}$ because the defense is not evolved until time $t_{i+1}$. Consider the example in Fig. \ref{fig:framework} (a), where the defense generations evolve at $t=0, 3$ and $4$, meaning $t_0=0$, $t_1=3$, $t_2=4$, $\D_0=\D_1=\D_2$, and $\D_4=\D_5=\D_6$. Therefore, the defender's GT is a random variable sampled by $\GT(\D,0)=t_1-t_0=3$ and $\GT(\D,1)=t_2-t_1=1$ in this toy example.

{\bf Attackers' GT, denoted by $\GT(\A)$:} 
Suppose the attack evolves at $t'_0=0,t'_1,\ldots,t'_k \leq T$, namely $\{t'_0,t'_1,\ldots,t'_k\}\subseteq [0,T]$, where notation $t'$ (rather than $t$) is meant to further highlight the perspective of the attacker's. Then, the attacker's GT, namely random variable $\GT(\A)$, is sampled by $\GT(\A,j)$'s, where
\begin{equation}
\GT(\A,j)=t'_{j+1}-t'_j ~~\text{for}~~ j=0,1,\ldots,k-1.
\end{equation}
This means that $\A_{{t'_j}+\Delta t} = \A_{t'_j}$ for any $\Delta t < t'_{j+1}-t'_j$ and that $\A_{t'_j+\GT(\A,j)} = \A_{t'_{j+1}} \neq \A_{t'_j}$.
Consider the example illustrated in Fig. \ref{fig:framework} (a), where the defense is evolved at $t'=0, 4$ and $6$. This means $t'_0=0$, $t'_1=4$, $t'_2=6$, $\A_0=\A_1=\A_2=\A_3$, and $\A_4=\A_5$. The attacker's GT is a random variable sampled by $\GT(\A,0)=t'_1-t'_0=4$ and $\GT(\A,1)=t'_2-t'_1=2$ in this toy example.

Summarizing the preceding discussion, we have:
\begin{definition}
	\label{definition:adaptation-latency}
	\emph{\bf (GT)} The defender's GT is defined as random variable $\GT(\D)$, sampled by:
	\begin{equation}
	\GT(\D,0)=t_1-t_0,
	\ldots, \GT(\D,\ell-1)=t_\ell-t_{\ell-1},
	\end{equation}
	where $t_0=0,t_1,\ldots,t_\ell\leq T$ are the sequence of points in time the defense evolves.
	The attacker's GT is defined as a random variable $\GT(\A)$, sampled by:
	\begin{equation}
	\GT(\A,0)=t'_1-t'_0,
	\ldots, \GT(\A,\ell-1)=t'_\ell-t'_{\ell-1},
	\end{equation}
	where $t'_0=0,t'_1,\ldots,t'_k\leq T$ are the sequence of time points to which the attack evolves.
\end{definition}

\subsubsection{\bf Effective-Generation-Time (EGT)} This metric considers the attack-defense generations as a whole by measuring the time to make \emph{effective} generations by an attacker or defender. Note that EGT is different from GT because the latter only focuses on timeliness. Moreover, GT may not be able to reveal a relationship with respect to the opponent's generations because not every generation is the result of adversarial action. 
Measuring EGT will allow a party to characterize in retrospect the effectiveness of its strategies over the time horizon. 

{\bf Defenders' EGT, denoted by $\EGT(\D)$:} Suppose defense generations evolve at $t_0=0,t_1,\ldots,t_\ell\leq T$, namely $\{t_0,t_1,\ldots,t_\ell\}\subseteq [0, T]$. The defender's EGT is a \emph{random variable}, denoted by $\EGT(\D)$, because the evolution of defense generations are stochastic in nature. The random variable $\EGT(\D)$ is sampled by $\EGT(\D,i)$'s for $i=0,\ldots,\ell-1$ such that $\D_{t_i+\EGT(\D,i)}$ is the nearest future generation that leads to a higher than $\D_{t_i}(\A_{t_i},M)$ defense effectiveness. Formally, $\EGT(\D,i)$ is defined as
\begin{equation}
\EGT(\D,i)=t_{i^*}-t_i
\end{equation}
when there exists some $t_{i^*}\in \{t_{i+1},\ldots,t_\ell\}$ such that
\begin{eqnarray}
\label{eq:condition-for-defender-effective-adaptation-1}
\D_{t_i+\Delta t}(\A_{t_i},M) \leq \D_{t_i}(\A_{t_i},M) \\ ~\text{for any}~0< \Delta t < \EGT(\D,i) \nonumber
\end{eqnarray}
and
\begin{equation}
\label{eq:condition-for-defender-effective-adaptation-2}
\D_{t_i+\EGT(\D,i)}(\A_{t_i},M) =\D_{t_{i^*}}(\A_{t_i},M)> \D_{t_i}(\A_{t_i},M);
\end{equation}
otherwise, we define $\EGT(\D,i)=\infty$, indicating that from the defender's perspective,
no further effective defense generation is made against attack $\A_{t_i}$ after $t_i$.

\begin{figure}[htbp!]
	\centering
	\subfigure[Defense generations at time $t=0, 3, 4$, where 
	$\GT(\D,0) = 3$ and $\EGT(\D,0) = 4$ as discussed in the text.]{\includegraphics[width=.44\textwidth]{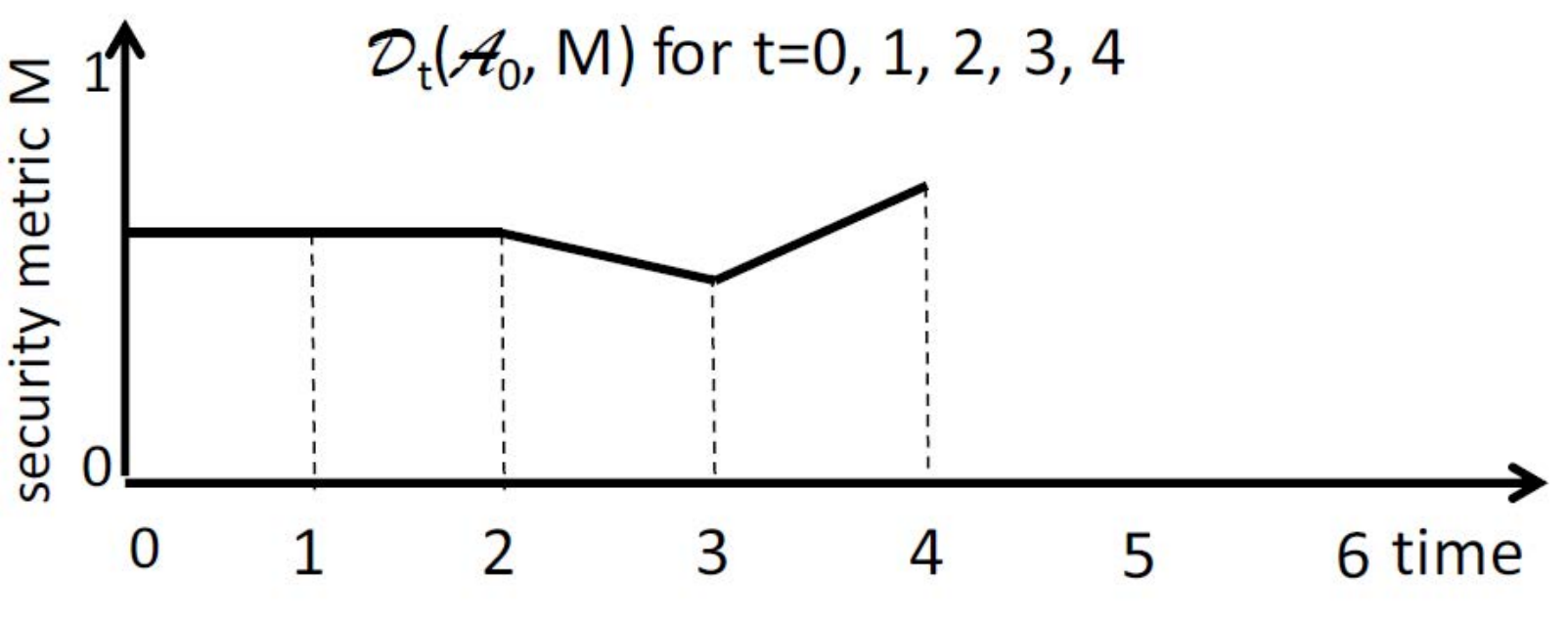}}
	\hfill
	\subfigure[Attack generations at $t=0, 4, 6$, where $\GT(\A,0)=4$ and $\EGT(\A,0) = 4$ as discussed in the text.]{\includegraphics[width=.44\textwidth]{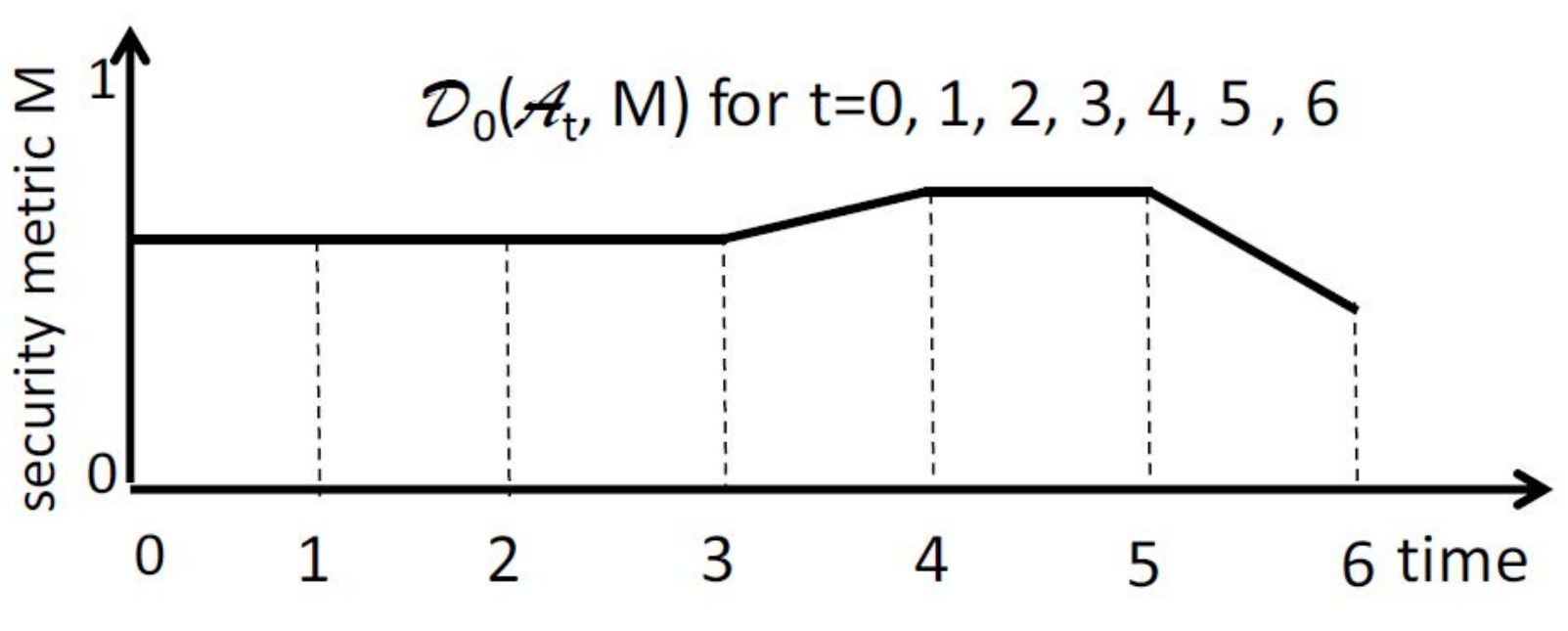}}
	\hfill
	\caption{GT (Generation-Time) vs. EGT (Effective-Generation-Time).}
	\label{fig:illustration}
\end{figure}

Let us continue to use the example in Fig. \ref{fig:framework}, where the three defense generations are respectively evolved at $t_0=0$, $t_1=3$, and $t_2=4<T=6$. Fig. \ref{fig:illustration} (a) describes the defense effectiveness of $\D_{t}(\A_0,M)$ for $t=0, 1, 2, 3, 4$. Since $\D_3(\A_0,M)<\D_0(\A_0,M)$, the defense generation at $t=3$ is \emph{not} effective. Since $\D_4(\A_0,M)>\D_0(\A_0,M)$, the defense generation at $t=4$ is effective. Therefore, we have $\EGT(\D,0)=t_2-t_0=4>\GT(\D,0)=3$. Moreover, suppose $\D_4(\A_3,M)>\D_3(\A_3,M)$,
meaning that the generation evolved at time $t_2=4$ is more effective than the previous generation evolved at time $t_1=3$. Then, we have $\EGT(\D_{t_1},1)=t_2-t_1=1$; otherwise, we have $\EGT(\D_{t_1},1)=\infty$, indicating that no more effective defense evolution is made against attack $\A_{t_1}$ after time $t_1=3$.

{\bf Attackers' EGT, denoted by $\EGT(\A)$:} Suppose the attack generation evolve at
$
t'_0=0, t'_1,\ldots,t'_k\leq T,
$
namely $\{t'_0,t'_1,\ldots,t'_k\}\subseteq [0,T]$. The attacker's EGT is defined as a \emph{random variable} $\EGT(\A)$ because the evolution events are stochastic.
The random variable $\EGT(\A)$ is sampled by
$
\EGT(\A,j) ~~\text{for}~~ j=0,\ldots,k-1
$
such that
$\D_{t'_j+\EGT(\D,j)}$ is the nearest-future defense generation that leads to defense effectiveness that is smaller than $\D_{t'_j}(\A_{t'_j},M)$ in terms of a metric $M\in \M$. Formally, $\EGT(\A,j)$ is defined as
\begin{equation}
\EGT(\A,j) = t'_{j^*}-t'_j
\end{equation}
when there exists $t'_{j^*}\in \{t'_{j+1},\ldots,t'_k\}$ such that
\begin{eqnarray}
\label{eq:condition-for-attacker-effective-adaptation-1}
\D_{t'_j}(\A_{t'_j+ \Delta t},M) \geq \D_{t'_j}(\A_{t'_j},M) \\ ~\text{for any}~ 0< \Delta t < \EGT(\A,j) \nonumber
\end{eqnarray}
and
\begin{equation}
\label{eq:condition-for-attacker-effective-adaptation-2}
\D_{t'_j}(\A_{t'_j +\EGT(\A,j)},M) =\D_{t'_{j}}(\A_{t'_{j^*}},M)< \D_{t'_j}(\A_{t'_j},M);
\end{equation}
otherwise, we define $\EGT(\A,j)=\infty$, meaning that no effective attack generation is evolved against $\D_{t'_j}$ at $t'_j$.

Let us continue to use the example described in Fig. \ref{fig:framework}, where the three attack generations are respectively evolved at time $t'_0=0$, $t'_1=4$, and $t'_2=T=6$.
Fig. \ref{fig:illustration}~(b) describes the defense effectiveness of $\D_{0}(\A_{t'})$ for $t'=0,\ldots,6$. Since $\D_0(\A_4,M)>\D_0(\A_0,M)$, the attack generation at time $t=4$ is not effective from the attacker's perspective. Since $\D_0(\A_6,M)<\D_0(\A_0,M)$, the attack generation at time $t=6$ to $\D_0$ is effective from the attacker's perspective. Therefore, we have $\EGT(\A,0)=t'_{2}-t'_0=6>\GT(\A,0)=4$.

Summarizing the preceding discussion, we have:
\begin{definition}
	\label{definition:effective-adaptation-latency}
	\emph{({\bf EGT})}
	Let $\M$ be the universe of metrics measuring static defense effectiveness discussed above. Suppose the defense generation evolve at $t_0=0,\ldots,t_\ell$ where $t_\ell \leq T$. The defender's EGT is defined as a random variable $\EGT(\D)$ sampled by
	the $\EGT(\D,i)$'s, where $i=0,1,\ldots,\ell-1$, that satisfy conditions in Eqs. \eqref{eq:condition-for-defender-effective-adaptation-1} and \eqref{eq:condition-for-defender-effective-adaptation-2}.
	
	Suppose the attack generations evolve at $t'_0=0,t'_1,\ldots,t'_k\leq T$.
	The attacker's EGT is defined as a random variable $\EGT(\A)$ sampled by
	the $\EGT(\A,j)$'s, where $j=0,1,\ldots,k-1$, that satisfy conditions in Eqs. \eqref{eq:condition-for-attacker-effective-adaptation-1} and \eqref{eq:condition-for-attacker-effective-adaptation-2}.
\end{definition}

{\bf Remark.} When comparing Definitions \ref{definition:adaptation-latency} and \ref{definition:effective-adaptation-latency}, we can derive:
\begin{equation}
\GT(\D,t) \leq \EGT(\D,t) ~~\text{and}~~ \GT(\A,t) \leq \EGT(\A,t)
\end{equation}
for any $t$.
In Definition \eqref{definition:effective-adaptation-latency}, it is possible that $\EGT(\D,i)=\infty$ for some $i\in [0,\ldots,\ell-1]$, indicating that the attack generation $\A_{t_i}$ occurred at time at $t_i$ is not addressed or countered by any later defense generation. Similarly, $\EGT(\A,j)=\infty$ for some $j\in[0,\ldots,k-1]$ indicates that the defense generation $\D_{t'_j}$ occurred at time $t'_j$ is not countered by the attacker in any later attack generation.

\subsubsection{\bf Triggering-Time (TT)}
This metric aims to answer the intuitive question ``which generations may have caused or \emph{triggered} which of the opponent's generations''. This would offer valuable insights into the opponent's operational process, especially a sense of \emph{responsiveness}. 
However, the measurement result does not necessarily represent the causal triggering of a given generation (e.g., a moving-target defense may not be causally related to any specific attack but may be used to increase the attacker's attack effort or cost).

\begin{figure}[htbp!]
	\centering
	\subfigure[Defense TT]{\includegraphics[width=.4\textwidth]{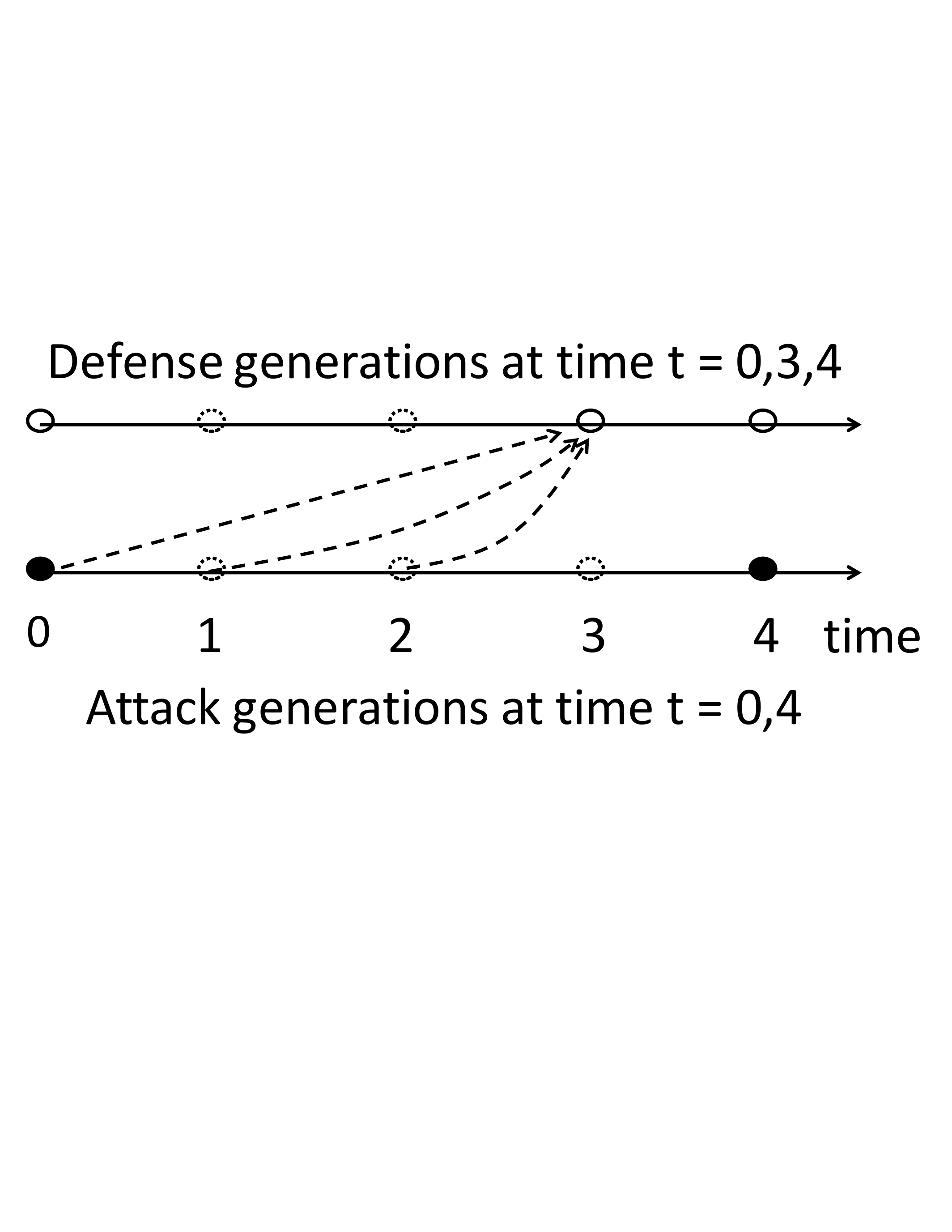}}\\
	\subfigure[Attack TT]{\includegraphics[width=.4\textwidth]{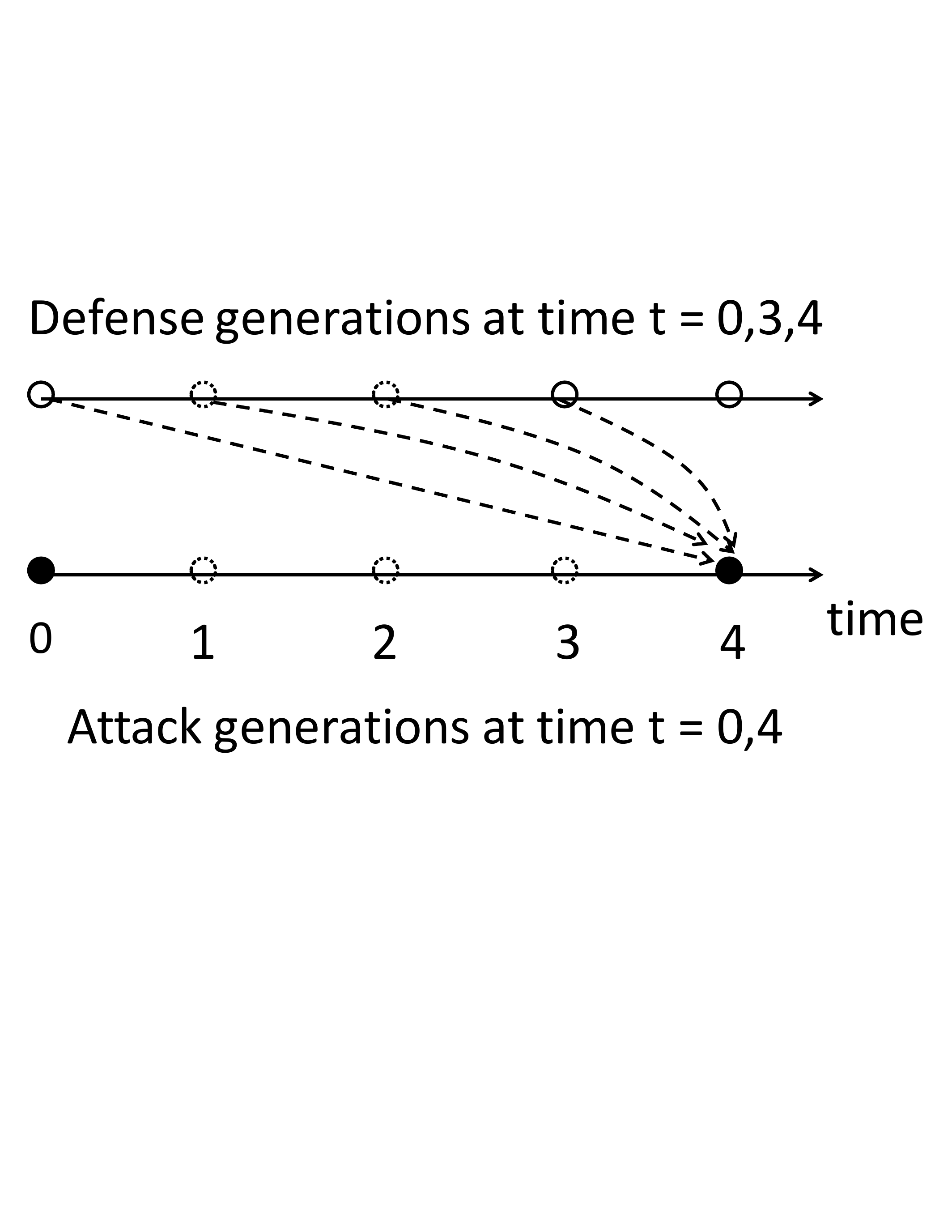}}
	\caption{TT (Triggering-Time) metric: A dashed arrow from generation $X$ to the opponent's generation $Y$ indicates that ``$X$ may have triggered $Y$.''}
	\label{fig:triggering-event}
\end{figure}

{\bf Defender's TT, denoted by $\TT(\D)$:} Fig.~\ref{fig:triggering-event} is obtained by splitting Fig.~\ref{fig:framework}~(a) into two pictures to explain how TT is measured. Recall that defense generations are evolved at time $t=0, 3, 4$, but here it suffices to consider only the two defense generations at $t=0$ and $t=3$ as an example. In Fig. \ref{fig:triggering-event}(a), the defense generation at $t=3$, namely $\D_3$, may be triggered by some of the attack generations $\A_0$, $\A_1$, and $\A_2$ that have been made by the attacker. We may define the \emph{triggering-event} of $\D_3$ as $\A_j$ for some $j\in[0,2]$ such that $\D_3(\A_j,M)$ has the greatest positive change in defense effectiveness when compared to $\D_0(\A_j,M)$, where $\D_0$ is considered because it represents the previous defense generation at $t=0$, and $j\in [0,2]$ is considered because $\A_0,\A_1$ and $\A_2$ represent the entire history of attacks in the time horizon. Suppose $\D_3(\A_j,M)-\D_0(\A_j,M)$ is maximized for some $j\in [0,2]$, suggesting that the defense generation may be triggered by the attack generation $\A_j$. This leads us to define the \emph{Triggering-Time} (TT) for defense generation $\D_3$ as $3-j$, which is a particular sample, denoted by $\TT(\D,3)$, of the random variable $\TT(\D)$ that is defined over the defense generations, except the $\D_0$ (because every sample needs to have a previous reference for comparison).

{\bf Attacker's TT, denoted by $\TT(\A)$:} For attack generations evolved at time $t=0$ and $4$, as shown in Fig.~\ref{fig:triggering-event}~(b), attack generation $\A_4$ may be \emph{triggered} by $\D_0$, $\D_1$, $\D_2$, or $\D_3$. We may define the \emph{triggering-event} of $\A_4$ as $\D_j$ for some $j\in [0,3]$ such that $\D_j(\A_4,M)$ has the greatest negative change in defense effectiveness when compared to $\D_j(\A_0,M)$, where $j\in[0,3]$ is considered because $\D_0,\D_1,\D_2$ \and $\D_3$ represent the history of defense generations up to time $t=4$, and $\A_0$ refers to the previous attack generation (prior to $\A_4$). Suppose  $\D_j(\A_4,M)-\D_j(\A_0,M)<0$ is minimized (i.e., maximized in its absolute value) at $j$, then we can define TT for attack generation $\A_4$ as $4-j$, which is a particular sample, denoted by $\TT(\A,4)$, of the random variable $\TT(\A)$ that is defined over the attack generations.

Summarizing the preceding discussion, we have:
\begin{definition}
	\emph{({\bf TT})}
	Suppose defense {generations} are evolved at $t_0=0, t_1,\ldots,t_\ell\leq T$ and attack generations are evolved at $t'_0=0,t'_1,\ldots,t'_k\leq T$. The \emph{triggering-event} for defense generation $\D_{t_i}$, where $i\in [1,\ell]$ is defined to be attack $\A_{t'}$ that leads to the greatest positive change in defense effectiveness relative to $\D_{t_{i-1}}$ in terms of metric $M\in \M$, namely
	\begin{eqnarray}
	\begin{gathered}
	\label{definition:TT-defense}
	t'={\argmax}_{0\leq t' < t_i} \D_{t_i}(\A_{t'},M) - \D_{t_{i-1}}(\A_{t'},M),\\
	\text{where}~~ \D_{t_i}(\A_{t'},M)>\D_{t_{i-1}}(\A_{t'},M). \nonumber
	\end{gathered}
	\end{eqnarray}
	If such $t'$ exists, we define $\TT(\D,i)=t_i-t'$; otherwise, we define $\TT(\D,i)=\infty$, meaning that the defense generation is not triggered by any past attack within the time horizon. The TT of defense generations is a \emph{random variable}, denoted by $\TT(\D)$, that is sampled by
	$\TT(\D,1),\ldots, \TT(\D,\ell)$.
	
	Similarly, the \emph{triggering-event} for attack generation $\A_{t'_j}$, where $j\in [1,k]$ is defined to be defense $\D_{t}$ that leads to the greatest negative change in defense effectiveness relative to $\A_{t'_{j-1}}$ in terms of metric $M\in \M$, namely
	\begin{eqnarray}
	t={\argmin}_{0\leq t < t'_j} \D_{t}(\A_{t'_j},M) - \D_t(\A_{t'_{j-1}},M), \\
	\text{where}~~\D_{t}(\A_{t'_j},M)< \D_t(\A_{t'_{j-1}},M). \nonumber
	\end{eqnarray}
	If such $t'$ exists, we define $\TT(\A,j)=t'_j-t$; otherwise, we define $\TT(\D,j)=\infty$, meaning that the attack generation is not triggered by any past defense within the time horizon. The TT of attack generations is a \emph{random variable}, denoted $\TT(\A)$, that is sampled by
	$\TT(\A,1),\ldots, \TT(\A,k)$.
\end{definition}

{\bf Remark.} 
The preceding definition of TT can be adapted in many flavors. In the preceding definition, we propose using the maximization of $\D_{t_i}(\A_{t'},M) - \D_{t_{i-1}}(\A_{t'},M)$ in Eq.~\eqref{definition:TT-defense} as the criterion for identifying triggering event. Alternatively, the definition can be adapted to maximize, for example, $\D_{t_i}(\A_{t'},M)$, meaning that defense $D_{t_i}$ is most effective against attack $\A(t')$. In the use case where both parties' evolution generations are completely known, the TT metric reflects the responsiveness of a party. In the case the opponent's evolution generations are not completely known (but the party's own evolution generations are naturally known), the TT metric can be used to identify, in retrospect, a party's evolution generation that may be the result of non-adversarial changes that have a security effect (e.g., new feature releases or patching of vulnerabilities). Such a retrospective security analysis is important because it helps the defender identify effective defense activities that would not be noticed by the defender otherwise. This is important because these possibly unconscious defense decisions and can offer insights into effective defense (e.g., best practice).

\subsubsection{\bf Lagging-Behind Time (LBT)}

This metric aims to measure how far one party is behind its opponent.  Although professionals have often said that the defender lags behind the attacker, we define this metric from both the defender's and attacker's perspectives because they could provide ways for proactive defenses ahead of actions by the attacker. 

\begin{figure}[htbp!]
	\centering
	\subfigure[$D_t(A_{t-1}, M)$]{\includegraphics[width=.44\textwidth]{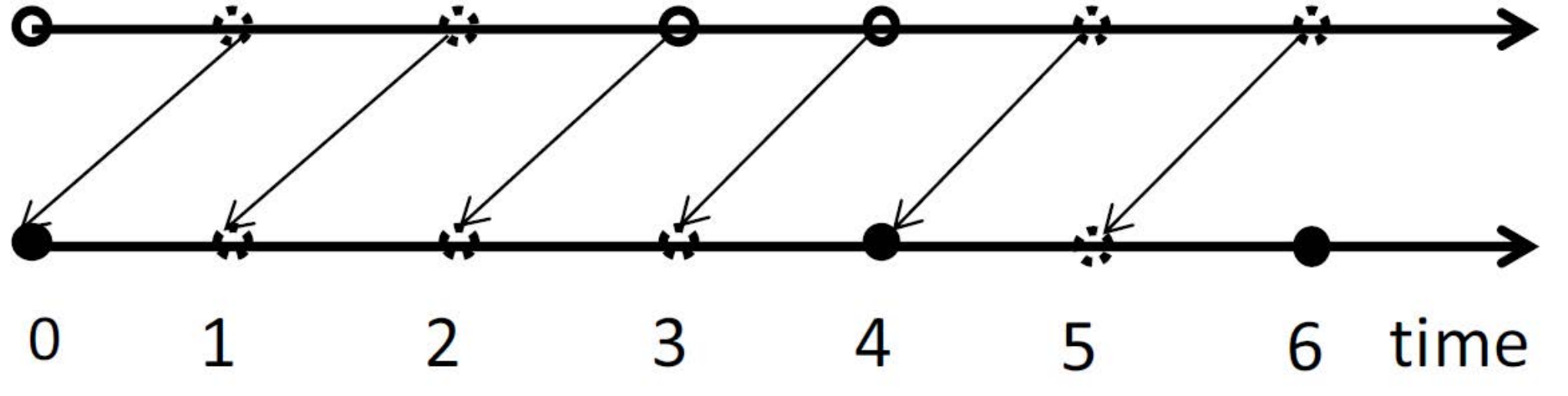}}
	\subfigure[$D_t(A_{t+1}, M)$]{\includegraphics[width=.44\textwidth]{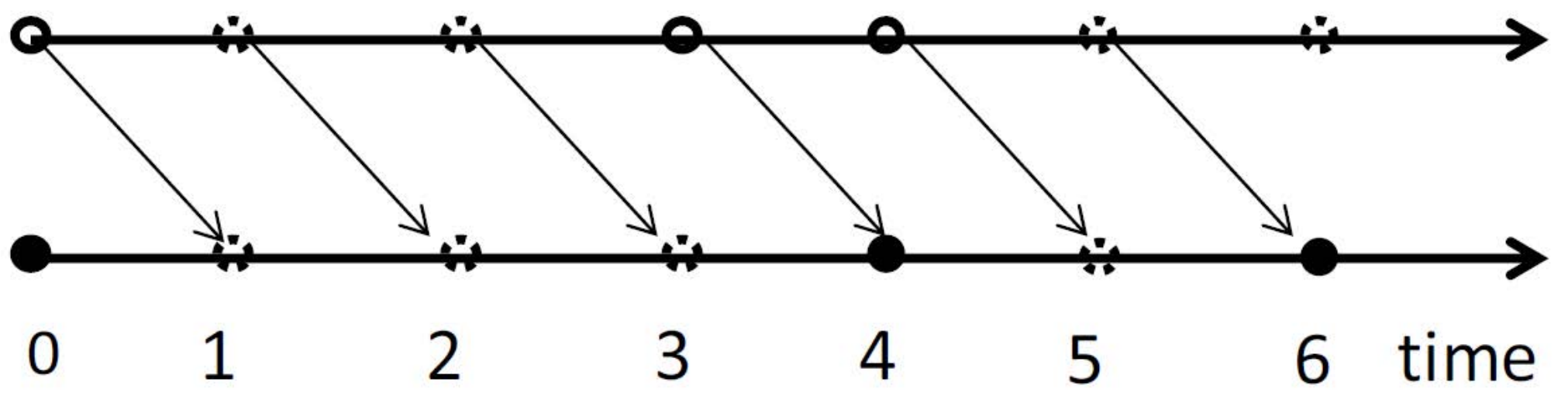}}
	\caption{Attacker and Defender LBT (Lagging-Behind Time).}
	\label{fig:lagging-behind}
\end{figure}

{\bf Defender's LBT, denoted by $\LBT(\D)$:}
For a security metric $M\in \M$ of interest, let $\varepsilon$, where $0\leq\varepsilon\leq 1$, represent the acceptable defense effectiveness. The LBT metric considers $\D_t(\A_{t-\lambda},M)$ for $\lambda =0,\ldots,T$ and $t\geq \lambda$. Fig.~\ref{fig:lagging-behind}~(a) illustrates $\D_t(\A_{t-\lambda},M)$ for $\lambda=1$. Then, we define $\LBT(\D)$ to be the \emph{minimum} $\lambda$ such that
\begin{equation}
\label{eq:defense-lagging-behind}
\D_t(\A_{t-\lambda},M)\geq \varepsilon ~~\text{for}~~ \lambda=0,\ldots,T~~\text{and}~~t\geq \lambda
\end{equation}
if such $\lambda$ exists; otherwise, we define $\LBT(\D)=-\infty$, meaning that defenses lag behind attacks at least for time $T$. In other words, we have
\begin{equation}
\label{eq:defense-lagging-behind-min}
\LBT(\D)=\min \{\lambda:\D_t(\A_{t-\lambda},M)\geq \varepsilon, \lambda=0,\ldots,T,~~t\geq \lambda\}.
\end{equation}
Note that $\varepsilon$
may vary from attack-defense settings (e.g., malware detection vs. intrusion detection). Note also that $\LBT(\D)=0$ means that the defender always keeps pace with the attacker.

{\bf Attacker's LBT, denoted by $\LBT(\A)$:} For a security metric $M\in \M$ of interest, recall that $\varepsilon$, where $0\leq\varepsilon\leq 1$, represents the acceptable defense effectiveness. This metric considers $\D_t(\A_{t+\lambda}, M)$ for $\lambda =0, \ldots, T$. Fig.~\ref{fig:lagging-behind}~(b) illustrates $\D_t(\A_{t+\lambda},M)$ for $\lambda=1$. Then, we define $\LBT(\A)$ to be the \emph{maximum} $\lambda$ such that
\begin{equation}
\label{eq:attack-lagging-behind}
\D_t(\A_{t+\lambda},M)\geq \varepsilon ~~\text{for}~~ \lambda=0,\ldots,T ~~\text{and}~~t\leq T-\lambda,
\end{equation}
if such $\lambda$ exists; otherwise, we define $\LBT(\A)=-\infty$, meaning that attacks lags behind defenses at least for time $T$.
In other words, we have
\begin{eqnarray}
\begin{gathered}
\label{eq:defense-lagging-behind-max}
\LBT(\A)=\max \{\lambda:\D_t(\A_{t+\lambda},M)\geq \varepsilon, \\
\text{where  } \lambda=0,\ldots,T,~~t\geq T-\lambda\}.
\end{gathered}
\end{eqnarray}
Note that $\LBT(\A)=0$ means that the attacker does not lag behind the defender.

Summarizing the preceding discussion, we have:
\begin{definition}
	\label{definition:LBT}
	$\LBT(\D)$ is defined by Eq.~\eqref{eq:defense-lagging-behind-min} with the minimum $\lambda$ that satisfies Eq.~\eqref{eq:defense-lagging-behind};
	$\LBT(\A)$ is defined by Eq.~\eqref{eq:defense-lagging-behind-max} with the maximum $\lambda$ that satisfies Eq.~\eqref{eq:attack-lagging-behind}.
\end{definition}

{\bf Remark.} Definition \ref{definition:LBT} can be relaxed by {adjusting} Eq.~\eqref{eq:defense-lagging-behind} such that:
\begin{equation}
\frac{1}{T-\lambda+1} \sum_{t=\lambda}^{T} \D_t(\A_{t-\lambda},M)\geq \varepsilon.
\end{equation}
This relaxation is to demand that the \emph{average} defense effectiveness is acceptable, rather than to demand that the defense effectiveness is \emph{always} acceptable. Note that $\LBT(\D)$ and $\LBT(\A)$ are ``dual'' to each other only in the sense that $\LBT(\D)$ looks backward in time while $\LBT(\A)$ looks forward in time.

\subsection{Effectiveness-Oriented Metrics}

\subsubsection{\bf Evolutionary Effectiveness (EE)}

This metric measures each generation with respect to a \emph{reference} generation. This is a random variable for each generation, sampled by the opponent's generations. 

\begin{definition}
	\label{def:EAE}
	\emph{({\bf EE})}
	Suppose defense generations are evolved at time $t_0=0,t_1,\ldots,t_\ell$ and attack generation are evolved at time $t'_0=0,t'_1,\ldots,t'_k$. Defender's EE is defined as a \emph{random variable}, denoted by $\EE(\D)$, which is sampled by $\EE(\D,j)$ for $j\in [1,k]$, where
	\begin{equation}
	\label{eq:EAE-D}
	\EE(\D,j)=\frac{1}{T+1} \sum_{t=0}^{T} [\D_t(\A_{t'_j},M)].
	\end{equation}
	With respect to a reference defense generation $\D_{t}$, the attacker's EE is defined as a \emph{random variable}, denoted by $\EE(\A)$, which is sampled by $\EE(\A,i)$ for $i\in [1,\ell]$, where
	\begin{equation}
	\EE(\A,i)=\frac{1}{T+1} \sum_{t'=0}^{T} [\D_{t_i}(\A_{t'},M)].
	\end{equation}
\end{definition}

{\bf Remark}. Definition~\ref{def:EAE} can be adapted by replacing  Eq.~\eqref{eq:EAE-D} with, for example, 
\begin{equation}
\EE(\D,j)=\frac{1}{T-t'_j+1} \sum_{t=t'_j}^{T} [\D_t(\A_{t'_j},M)].
\end{equation}

\subsubsection{\bf Relative-Generational-Impact (RGI)}



\begin{figure}[htbp!]
	\centering
	\includegraphics[width=.46\textwidth]{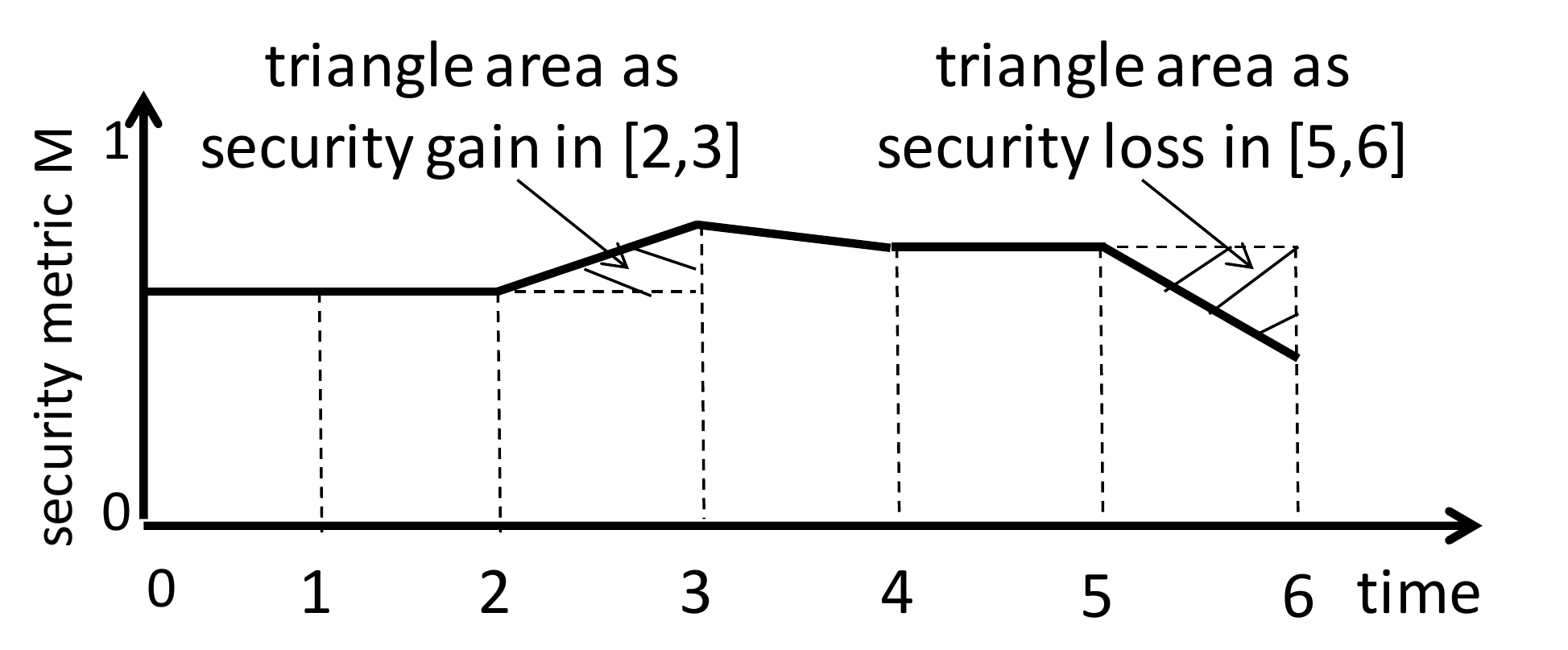}
	\caption{$\D_t(\A_t,M)$ over $[0,T]$: Defense generations are evolved at time $t=0, 3, 4$ while attack generations are evolved at time $t=0, 4, 6$.}
	\label{fig:adaptation-agility-illustration}
\end{figure}

As illustrated in Fig.~\ref{fig:adaptation-agility-illustration}, we propose comparing $\D_t(\A_t,M)$ and $\D_{t-1}(\A_{t-1}, M)$ for $t=1, \ldots, T$. At a specific point in time $t\in [1, T]$, there are three possible scenarios:
\begin{itemize}
	\item[(a)] When $\D_t(\A_t,M)=\D_{t-1}(\A_{t-1},M)$, the attacker's maneuver and the defender's maneuver at $t$ are equal.
	\item[(b)] When $\D_t(\A_t,M)>\D_{t-1}(\A_{t-1},M)$, the defender is out-maneuvering the attacker at $t$.
	\item[(c)] When $\D_t(\A_t,M)<\D_{t-1}(\A_{t-1},M)$, the attacker is out-maneuvering the defender at $t$.
\end{itemize}

\begin{definition}
	\emph{({\bf RGI})} Defender's RGI is a random variable, denoted by $\RGI(\D)$ and sampled by $\RGI(\D,t)$ for $t=1, \ldots, T$, where
	\begin{equation}
	\RGI(t)=\D_t(\A_t,M)-\D_{t-1}(\A_{t-1},M).
	\end{equation}
\end{definition}

Note that unlike the metrics mentioned above, we omit an attacker's RGI because it would hold that 
$\RGI(\A,t)=-\RGI(\D,t)$ for $t=1,\ldots,T$.

\subsubsection{\bf Aggregated-Generational-Impact (AGI)}
This metric aims to measure the overall security gained by the defender over time horizon $[0,T]$. We propose measuring the \emph{security gain} over time interval $[t_{i-1},t_i]$, denoted by $\mathcal{G}_s(i)$, for $i=1, \ldots, T$. As Fig.~\ref{fig:adaptation-agility-illustration} describes, we assume that a straight-line is used to link $(t_{i-1}, \D_{t_i-1}(\A_{t_i-1},M))$ and $(t_{i}, \D_{t_i}(\A_{t_i},M))$.
Then, $\mathcal{G}_s(i)$ is defined as the area of the triangle that depends on the sign of $[\D_{t_i}(\A_{t_i},M)-\D_{t_i-1}(\A_{t_i-1},M)]$. More specifically, we define
\begin{eqnarray}
\label{eq:security-gain-local}
&&\mathcal{G}_s(i) \nonumber\\
&=&
\left\{
\begin{array}{ll}
\frac{1}{2}[\D_{t_i}(\A_{t_i},M)-\D_{t_i-1}(\A_{t_i-1},M)], \\
~~~~~~~~~~ \text{if}~\D_{t_i}(\A_{t_i},M)>\D_{t_i-1}(\A_{t_i-1},M) \\
-\frac{1}{2}[\D_{t_i}(\A_{t_i},M)-\D_{t_i-1}(\A_{t_i-1},M)], \\
~~~~~~~~~~ \text{if}~\D_{t_i}(\A_{t_i},M)<\D_{t_i-1}(\A_{t_i-1},M) \\
0 ~~~~~~~~~\text{if}~\D_{t_i}(\A_{t_i},M)=\D_{t_i-1}(\A_{t_i-1},M),
\end{array}
\right.
\end{eqnarray}
where $t_i-t_{i-1}=1$. This leads to:
\begin{definition}
	\label{definition:global-stability}
	\emph{({\bf AGI})} Defender's AGI over $[0,T]$, denoted by $\AGI(\D)$, is defined by
	\begin{equation}
	\AGI(\D)=\frac{1}{T} \sum_{i=1}^T \mathcal{G}_s(i).
	\end{equation}
\end{definition}
{An} attacker's AGI is omitted because {its definition is trivial} as $\AGI(\A)=-\AGI(\D)$.

\begin{figure}[htbp!]
	\centering
	\includegraphics[width=.46\textwidth]{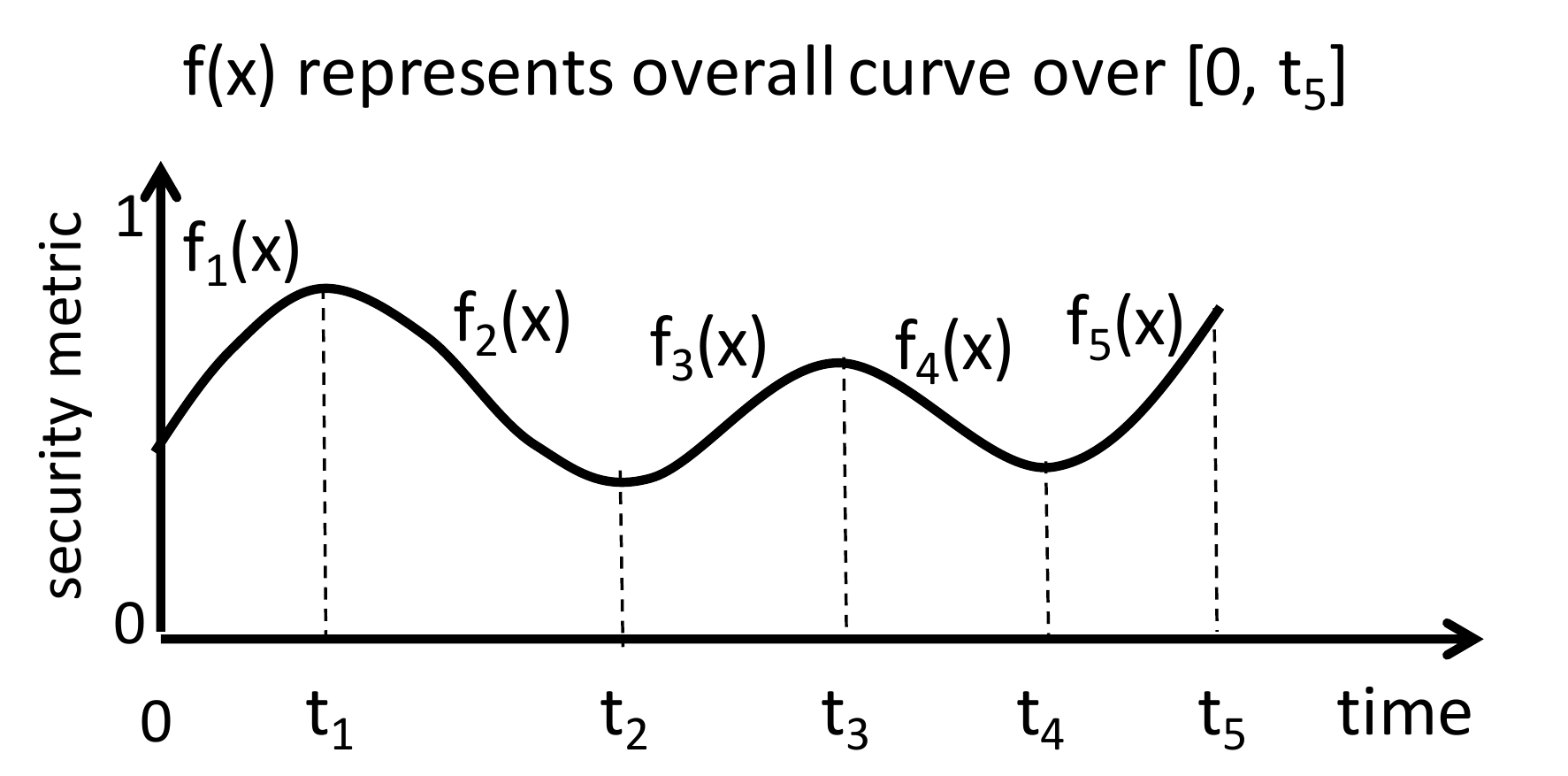}
	\caption{Justification on using {an} area to define \emph{security gain} $\mathcal{G}_s(i)$.}
	\label{fig:area-justification}
\end{figure}

{\bf Remark.} Note that the length of each time interval is $t_i-t_{i-1}=1$ for $i=1, \ldots, T$ because the time horizon is defined as $t=0, 1, \ldots, T$.
As such, one may observe that $\mathcal{G}_s(i)$ as shown in Eq.~\eqref{eq:security-gain-local} does not have to be interpreted as using the areas of triangles corresponding to individual time intervals; instead, it can be interpreted directly as $\sum_{i=1}^T \mathcal{G}_s(i)$ while ignoring the constant $\frac{1}{2}$.
Nevertheless, it has two advantages to use areas of the triangles to define \emph{security gain} $\mathcal{G}_s(i)$: (i) This definition remains equally applicable when the time intervals do \emph{not} have the same length, as illustrated in Fig.~\ref{fig:area-justification}; and (ii) when there is a need to interpolate a continuous and smooth curve of $\D_t(\A_t,M)$ over $[0,T]$, such as the curve $f(x)$ in Fig.~\ref{fig:area-justification}, we can divide the curve $f(x)$ into $z$ segments such that within each segment, $f(x)$ is strictly monotonic. In the example of Fig.~\ref{fig:area-justification}, $f(x)$ is divided into 5 segments, $f_1(x),\ldots,f_z(x)$. Accordingly, we can extend Definition~\ref{definition:global-stability} to define AGI over $[0,T]$ as:
\begin{equation}
\AGI(T)=\frac{1}{T}\sum_{i=1}^z \mathcal{G}_s(i)
\end{equation}
where
\begin{eqnarray}
&&\mathcal{G}_s(i)  \\
&=&\left\{
\begin{array}{ll}
\int_{{t_{i-1}}}^{t_i}f_i(x)dx - (t_i-t_{i-1})f_i(t_{i-1})  &\text{if}~~ f'_i(x)> 0 \nonumber \\
(t_i-t_{i-1})f_i(t_{i-1}) - \int_{{t_{i-1}}}^{t_i}f_i(x)dx  &\text{if}~~ f'_i(x)< 0 \nonumber \\
0 &\text{if}~~ f'_i(x)= 0. \nonumber
\end{array}
\right.
\end{eqnarray}

\section{Discussion} \label{sec:discussion}

In this section, we discuss use cases of the proposed metric framework. 
Since our goal is to help defenders, the discussion below will be from a defender's point of view. We focus on two issues: the amount of attack evolution generations that are observed by the defender; and the number of attackers vs. the number of defenders. 

\subsection{Use Case with respect to the Amount of Attack Generations Observed}

We differentiate two scenarios: \emph{all} vs. \emph{some or no} attack evolution generations being observed.

\subsubsection{Use Cases When All Evolution Generations Are Observed}

This is the ideal case and is possible when considering (for example) white-hat attack defense experiments over a period of time. 
This use case is also possible for \emph{retrospective attack-defense analysis}. 
This can happen because some attacks (e.g., new attacks or even zero-day attacks) that take place at time $t$ may not be recognized until time $t'$, where $t<t'$. If all cyber activities (e.g., network traffic and host execution) are properly recorded, the metrics defined above can be used to measure attack and defense evolution \emph{in retrospect}. The measurement results tell a defender about its agility and may lead to insights into explaining why the defense failed and how the failure may be fixed in the future. 

\subsubsection{Use Case When Some or No Evolution Generations Are Observed}

In real-world cyber attack-defense practice, the defender may only observe some or no attack evolution generations. 
In this case, the defender can identify ``probable'' attack evolution generations as follows. First, the defender treats each attack $\A_0,\ldots,\A_T$ at time $t\in [0,T]$ as an evolution generation, despite that some attacks are not evolution generations. 
Then, the defender can identify the attacks that disrupt the existing defense most as an approximation of attack evolution generations (i.e., ``probable'' generations as an approximation to the unknown ``ground-truth'' generations). For example, this can be done as follows: Given a threshold $\tau$ where $0<\tau<1$, attack $\A_{t'}$ can be treated as an evolution generation if there exists $t$, where $0\leq t < t'$, such that
\begin{equation}
\D_{t}(\A_{t},M) - \D_{t}(\A_{t'},M) > \tau.
\end{equation}
As a result, the defender can identify the probable generations and then use them measure the metrics proposed in the paper. 

From a conceptual point of view, the preceding use case is reminiscent of the ground truth in supervised machine learning. In principle, the training data in supervised machine learning should be 100\% accurate or correct (e.g., 0\% false-positives and 0\% false-negatives). In practice, this is hard to achieve. Still, supervised machine learning is useful and successful even if the ground truth of the training data is \emph{not} guaranteed. In this sense, the preceding method we propose using to identify attack evolution generations approximately might be as useful as in the case of supervised machine learning with an approximate ground-truth. 



\subsection{Use Cases with respect to the Number of Attackers vs. the Number of Defenders}
\label{sec:usefulness-of-the-framework}

There are 4 scenarios: one attacker against one defender; one attacker against multiple defenders; multiple attackers against one defender; and multiple attackers against multiple defenders. Since the description in Section \ref{sec:framework} focused on the scenario of one attacker against one defender, in what follows we discuss how the metrics framework can be used in the other three scenarios. 

\subsubsection{Use Case with One Attacker against Multiple Defenders}

This scenario is interesting when evaluating the collective effectiveness and failures of multiple defenders. From a defender's point of view, this is a straightforward extension to the preceding ``one attacker against one defender'' case because the defense generations that are evolved by each defender is known. In order to measure the collective defense effectiveness and failures, we can treat the collection of defenders as a single \emph{virtual} defender.

\subsubsection{Use Case with Multiple Attackers against One Defender}

This scenario is perhaps what happens in the real world where a defender (of an enterprise) needs to cope with multiple attackers. In this scenario, the multiple attackers can be represented by a single \emph{virtual} attacker. This makes sense when the attackers are coordinated because the coordinator can be seen as the attacker (while noting that this insight has been widely used in cryptographic models). This treatment also makes sense even if the attackers are not coordinated with each other because in the real world each defender (of an enterprise network, for example) is indeed likely dealing with multiple attackers. In this case, 
the attacks waged by different attackers can be superimposed over each other, leading to attack generations that are a superset of the generations that are evolved by individual attackers. As a result, the metrics framework is equally applicable in this scenario. 
This generality of the framework can be attributed to the fact that agility is about the attacks rather than the identities of the attackers.

We stress that the preceding discussion does not mean that effort should not be made to distinguish the attackers. This is because there is a spectrum of situations: at one end of the spectrum, the defender cannot tell the attackers apart; at the other end of the same spectrum, the defender can tell all of the attackers apart. In the former case, the defender has to treat all of the attackers as a single entity or coordinated one. In the latter case, the defender can measure cyber agility with respect to each recognized attacker, which allows the defender to measure which attacker(s) are more agile than the other attackers and therefore possibly prioritize defense resources against these more agile attackers. Therefore, the defender should always strive to distinguish the attacks waged or coordinated by different attackers.  In any case, our metrics framework is equally applicable.



\subsubsection{Use Case with Multiple Attackers against Multiple Defenders}

Similarly, this case can be seen as a simple extension to the case of ``one attacker against multiple defenders'' or the case of ``multiple attackers against one defenders.''

\section{Case Study} \label{sec:case-study}
\begin{figure*}[htbp!]
	\centering
	\subfigure[Defender's GT and EGT]{\includegraphics[width=.315\textwidth]{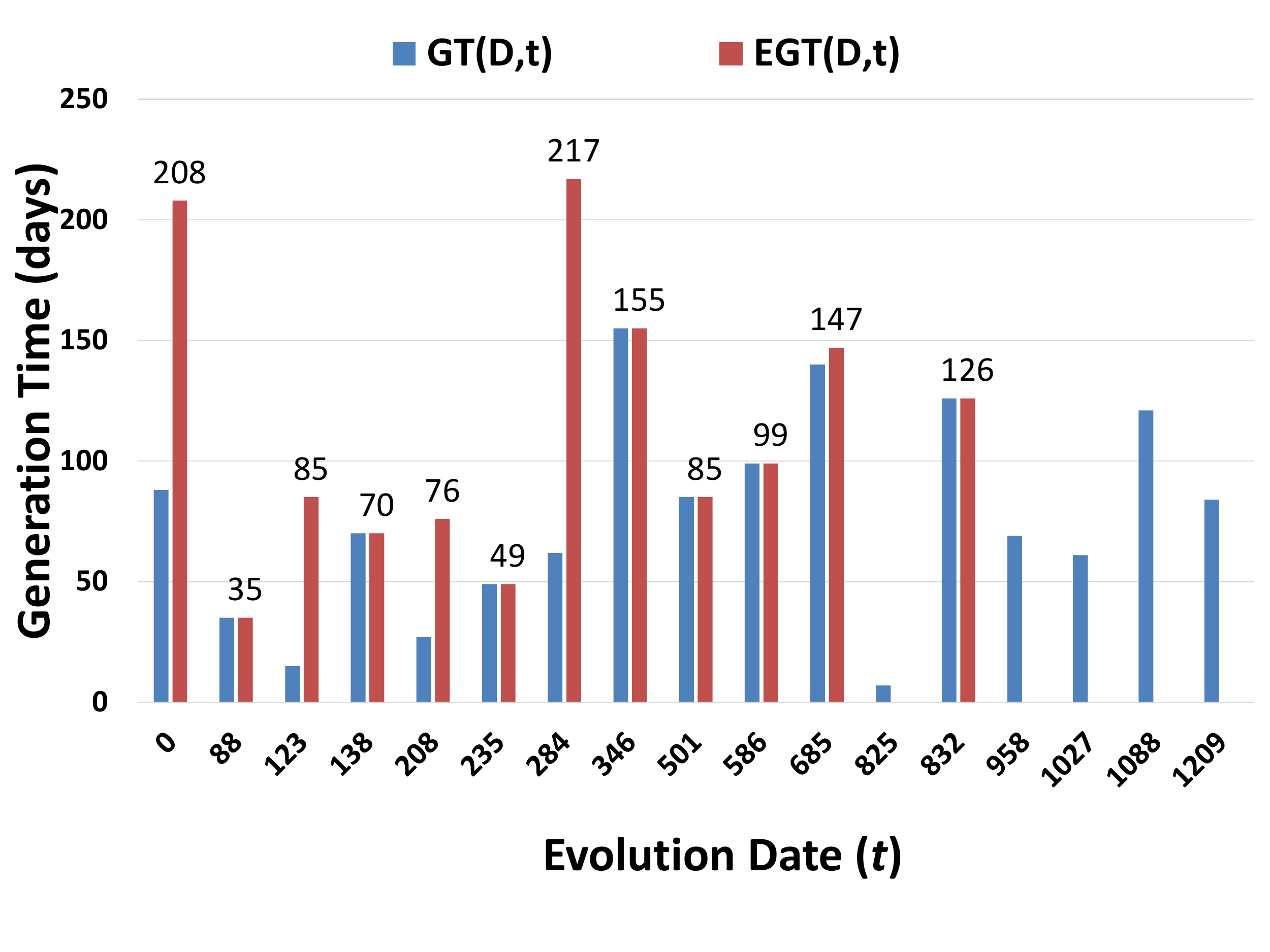} \label{hp_al}}
	\subfigure[Defender's TT]{\includegraphics[width=.315\textwidth]{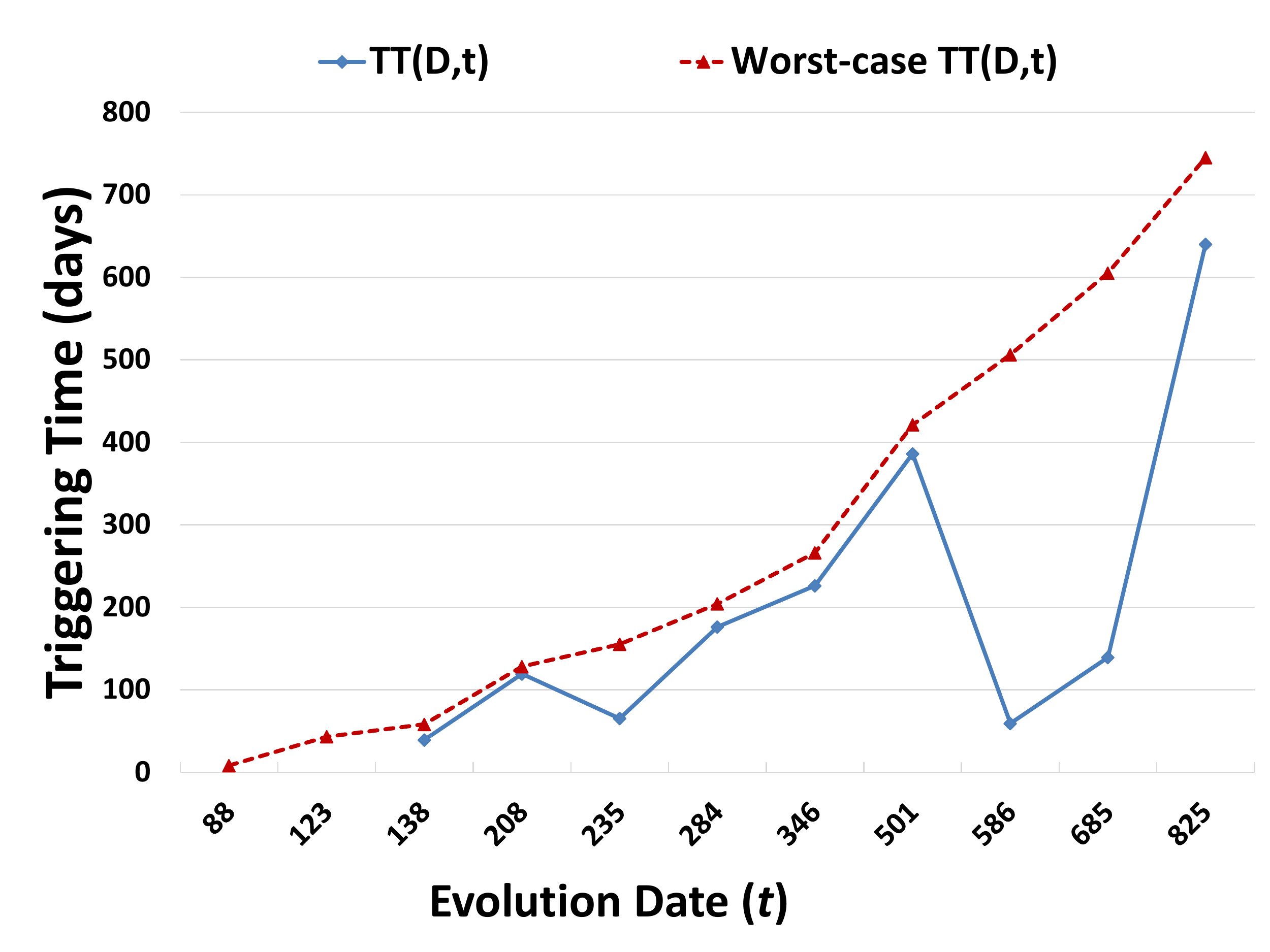} \label{hp_tt}}
	\subfigure[Defender's RGI]{\includegraphics[width=.315\textwidth]{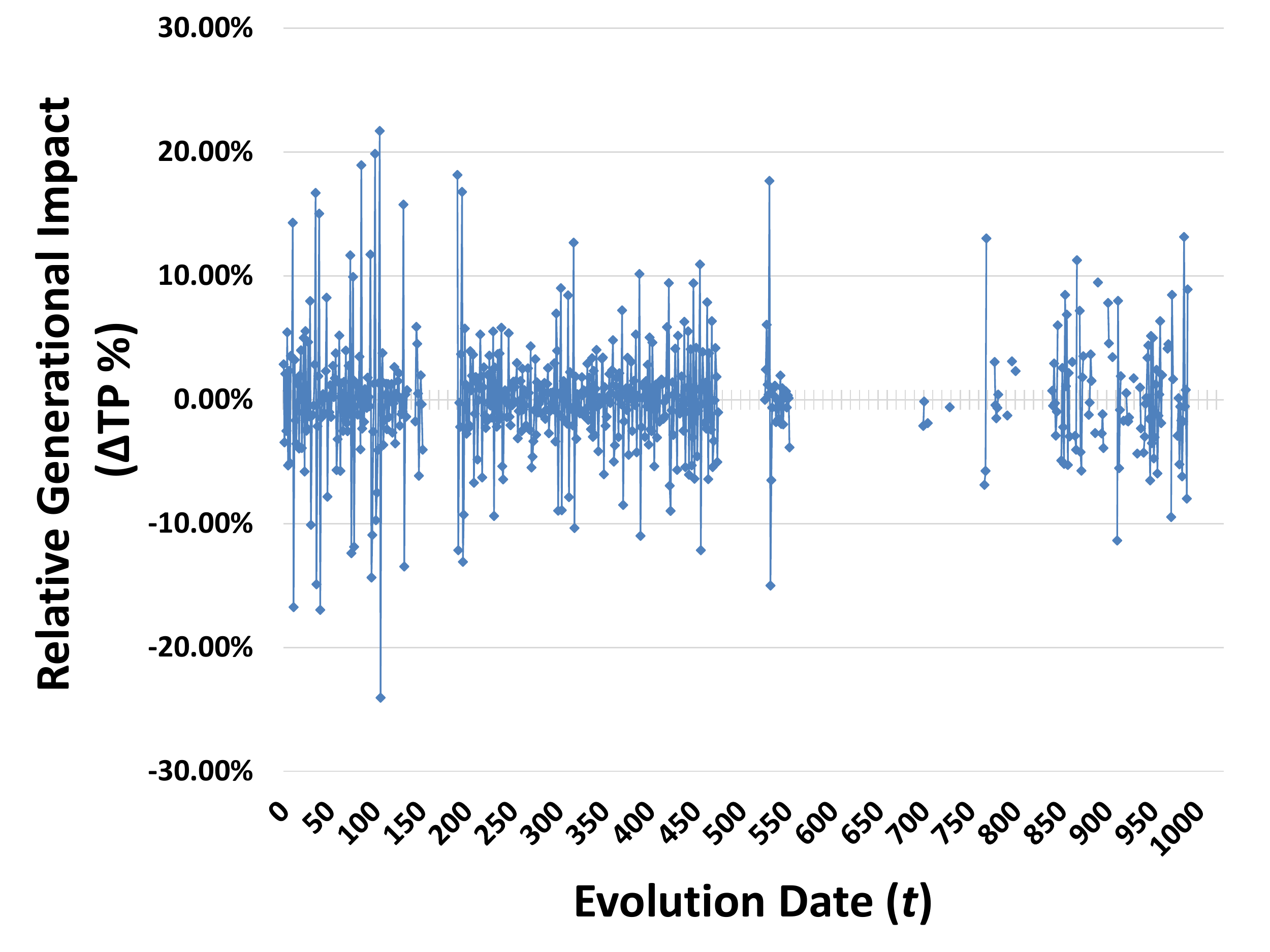} \label{hp_tae}}
	\caption{Evolution metrics measured for the defender (i.e., Snort) with the Honeypot dataset ($x$-axis unit: day).}
	\label{fig:Honeypot-Latency}
\end{figure*}

\begin{figure*}[htbp!]
	\centering
	\subfigure[Defender's GT and EGT]{\includegraphics[width=.32\textwidth]{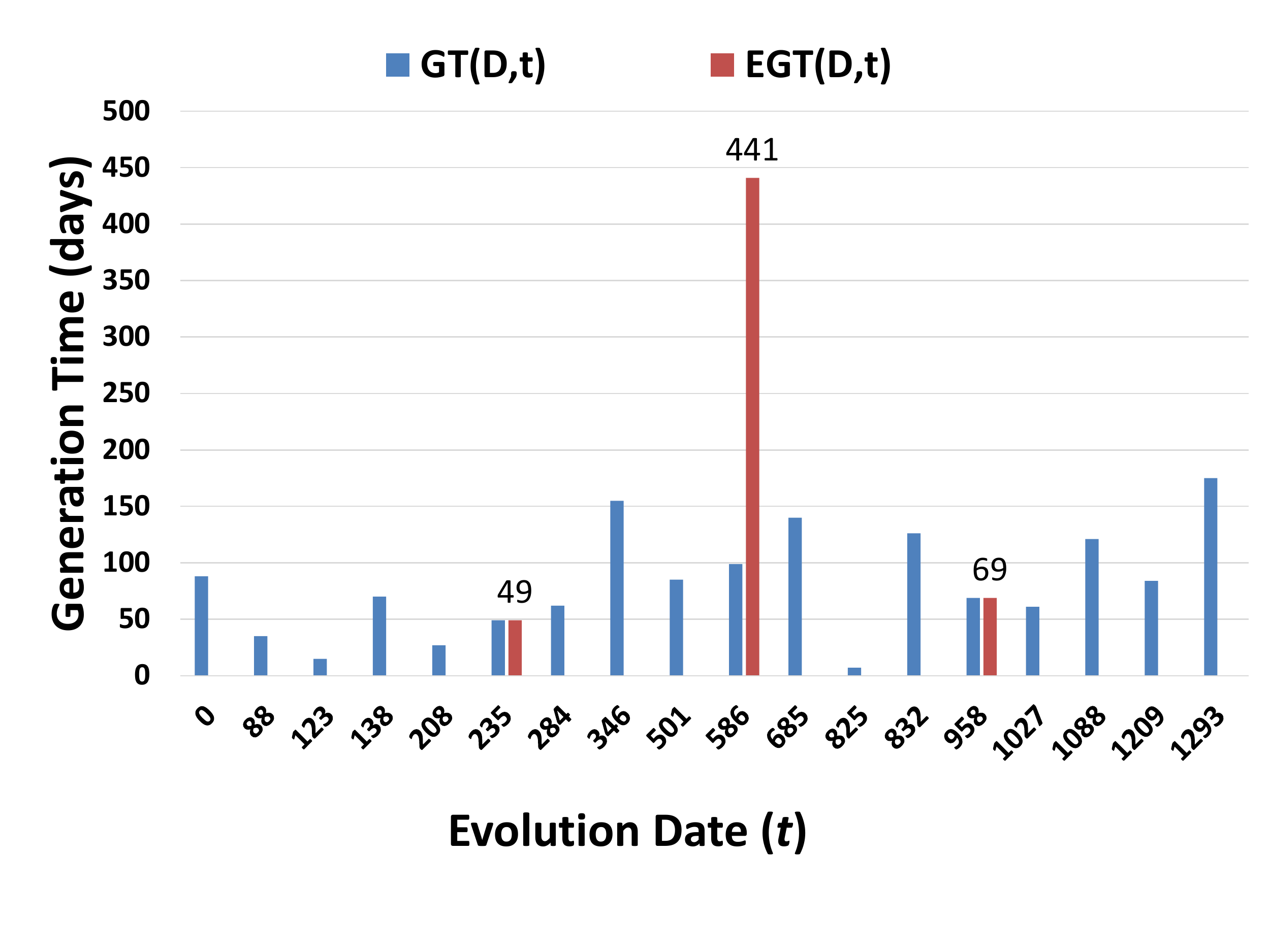} \label{df_al_d}}
	\subfigure[Defender's LBT]{\includegraphics[width=.32\textwidth]{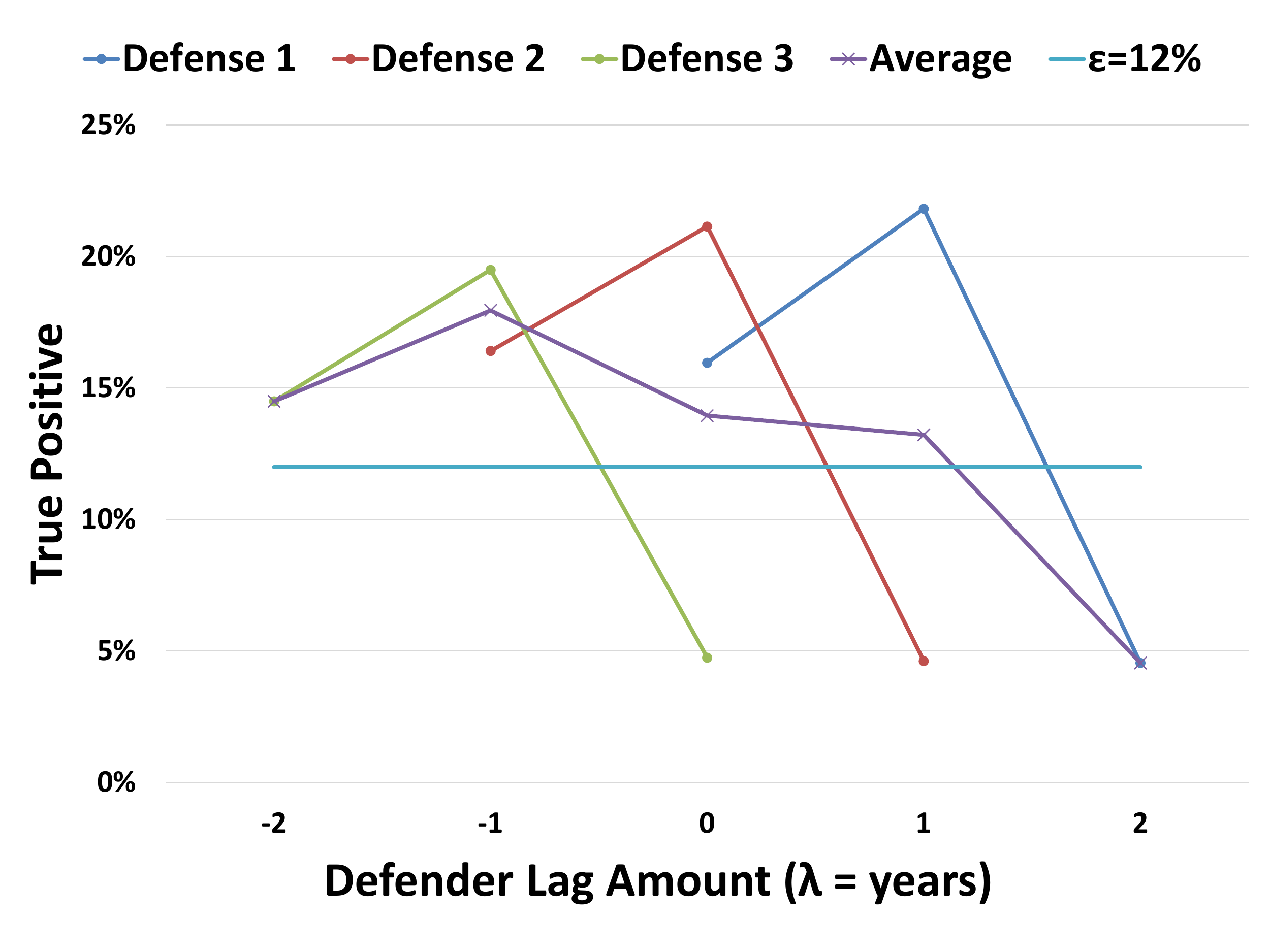} \label{df_lbt_d}}
	\subfigure[Defender's EE]{\includegraphics[width=.32\textwidth]{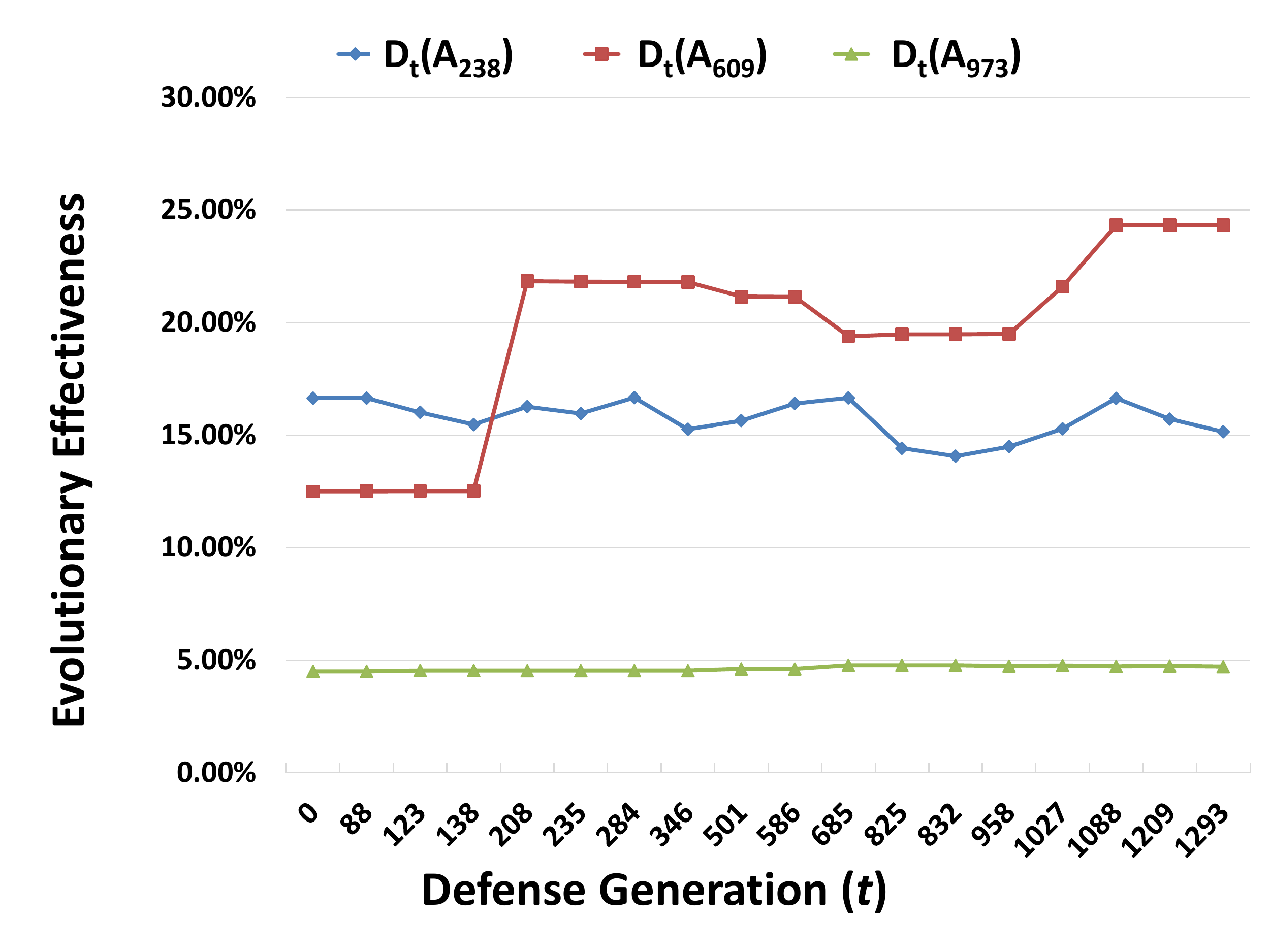} \label{df_eae_d}}
	\caption{Evolution metrics measured for the defender (i.e., Snort) with the DEFCON dataset ($x$-axis unit: day).}
	\label{fig:DF-Latency-Defender}
\end{figure*}

In this section, we show the results from our case study by applying the proposed metrics to two real datasets.

\subsection{Experimental Setup}

\subsubsection{Defense Tool} The case study is based on replaying network traffic, which contains attacks, against the Snort intrusion detection system~\cite{snortTool}.
We used six versions of Snort (v2.9.4 - v.9.8) released for Dec. 2012 - Dec. 2016. For each version (e.g., v2.9.4), there are sub-versions (e.g., v2.9.4.1). Each subversion is counted as a defense generation, leading to 18 versions or defense generations, denoted by $\D_{t_0},\ldots,\D_{t_{17}}$. These 18 generations are made for 1,294 days, meaning that the time horizon for the defender is $[t_0, t_{17}]=[1,1294]$. These Snort versions are tested on Virtual Machines (VMs) using the Ubuntu 14 operating system after making few changes on the default settings in the {\tt snort.conf} file to best fit for this experiment.

\subsubsection{Datasets} The 18 versions of Snort are tested against the following two datasets:
\begin{itemize}
	\item {\bf Honeypots dataset:} This dataset was collected at a low-interaction honeypot of approximately 7,000 IP addresses. This dataset contains traffic spanning Feb. 2013 - Dec. 2015.  The low-interaction honeypot consists of programs, including Dionaea, Mwcollector, Amun, and Nepenthes. These programs simulate services (e.g., SMB, NetBIOS, HTTP, MySQL, and SSH) to some extent. The dataset {was} collected over 1,029 days, with a time horizon for the attacker  $[t'_0, t'_{17}]=[80,1029]$. This is important because many datasets do not include such a length, especially in continuous (or mostly continuous) collections. Due to the absence of well-defined attack generations, we treat the attacks at each time unit (days) as an attack generation. This changes the measurement of some metrics, namely LBT and EE. There are some missing data during shorter time periods within this time horizon because the data collection system was occasionally shutdown due to various reasons. Although the dataset is not ideal because the attacker's and defender's time horizons are not exactly the same, which often happens in practice, the metrics framework can be appropriately applied by adjusting a different time span.
	
	\item {\bf DEFCON CTF dataset:} DEFCON is one of the world's premiere hacker-type conferences. We use the publicly available $pcap$ files collected from DEFCON 21 (2013), DEFCON 22 (2014) and DEFCON 23 (2015), each of which corresponds to a single day. This dataset has well defined attack generations (one per year). Putting into the terms of the metrics framework, the time horizon of the attacker's generations are at $t'_0=238$, $t'_1=609$, and $t'_2=973$. The dataset consists of $pcaps$ from 20 teams.  We randomly selected a single team's data for our experiment, and followed the team through all three DEFCON CTFs. Although DEFCON datasets are available for a number of years, we only consider these three because these three years correspond to the period of time during which Snort was considered as discussed above. It is critical that these years are distinct, because we are able to clearly see the difference between attack generations; this is not the case in the aforementioned honeypot dataset.
\end{itemize}

In our experiments, we replay the network traffic against Snort, log the alerts generated by Snort's preprocessor and detection engine, and calculate the true-positive rate, $tp$. This is possible because the datasets are composed of malicious traffic. In terms of the notations used above, the static metric $M$ used in the notation $\D_t(\A_{t'}, M)$ is the true-positive rate $tp$.
It is worth mentioning that as in any data-driven study, the insights derived from a specific dataset may not be arbitrarily generalized. When not all attack generations are observed, the resulting insights may be specific to the specification of generations used in the analysis in question.

\subsection{Case Study with the Honeypot Dataset}
Fig.~\ref{fig:Honeypot-Latency} plots the measured evolution metrics for the Honeypot dataset. Recall that the defender's time horizon is $[t_0, t_{17}]=[1,1294]$ and the attacker's time horizon is $[t'_0, t'_{17}]=[80,1108]$. This means that there are missing attack generations in the defender's time horizon, for which some metrics cannot be measured. We measure the proposed metrics by focusing on a defender's point of view. To be specific, we do not consider the lagging-behind time (LBT) and evolutionary-effectiveness (EE) metrics because these require both attack and defense generations to be well defined. The remainder of the metrics are well defined, given the constraints of this dataset.

{\bf Analysis on Fig.~\ref{fig:Honeypot-Latency}~(a):} This figure plots the sample of a defender's Generation Time, $\GT(\D)$, and the sample of the defender's Effective GT, $\EGT(\D)$. Note that some blue bars representing $\GT(\D)$ are not accompanied by red bars representing $\EGT(\D)$ due to the missing attack traffic data for the corresponding dates. Graphing both $\GT(\D)$ and $\EGT(\D)$ together allows us to see any disparity between these two metrics. We observe that the mean ratio of $\EGT(\D)$ to $\GT(\D)$ is 3.00. This implies that effective generations evolve 3x slower than normal generations. However, when we investigate individual generations, some defense generations are less responsive than others. For example, attack generation $\A_0$ is not effectively responded to by a defense generation until 208 days, as shown in $\EGT(\D,0)=208$.

{\bf Analysis on Fig.~\ref{fig:Honeypot-Latency}~(b) :} This figure shows the defender's Triggering-Time (TT) metric. The red curve indicates the {\em hypothetical} worst-case scenario in that every defense generation is triggered by the very first attack, $\A_{t_0}=\A_{80}$. The blue curve corresponds to $\TT(\D,t)$ as sampled at $t = 88, 123, 138, 208, 235, 284, 346, 501, 586, 685$, and $825$. Note that $\TT(\D,88)$ and $\TT(\D,123)$ are not shown because they are not obtainable from the dataset as they reach $\infty$, based on their definition. Note that $\TT(\D,t)$ for $t>825$ is not plotted because of missing attack traffic data at $t=825$. We observe that most defense generations respond to relatively old attacks, except $\TT(\D,235)=65$ and $\TT(\D,586)=59$, indicating that these defense generations respond to attacks that are almost two months old.

{\bf Analysis on Fig.~\ref{fig:Honeypot-Latency}~(c):} This figure plots the RGI metric, where there are some $t$'s at which $\RGI(\D,t)$ is missing because of missing attack traffic. From Fig.~\ref{fig:Honeypot-Latency}~(c), we can notice the true-positive rate is observed with a typical range of $\pm 10\%$. The aggregated generational effectiveness (AGI) of the defender (i.e., Snort) is very poor at $0.01\%$, implying little significant evolution relative to the attacker during the given times.

{\bf Results Analysis:}
Our results show that Snort has a history of being responsive to attacks in evolving its defense in a timely manner. However, the attackers also evolved, offsetting the previous defense gains. This explains why $\AGI(\D)\approx 0.01\%$, indicating that the Snort community is in a stalemate with the attacker. We notice that the static defense effectiveness metric, $\D_t(\A_t,M)$, is low with a mean of $8.11\%$. The cause of the low effectiveness is the low-interaction nature of the honeypot, which makes it not as semantically rich as we would like.

\ignore{
	\begin{figure*}[htbp!]
		\centering
		\subfloat[Defender's GT and EGT]{
			\includegraphics[width=.33\textwidth]{figures/fig11-al-eal-defcon.pdf} \label{df_al_d}
		}
		\subfloat[Defender's TT]{
			\includegraphics[width=.33\textwidth]{figures/fig13-tt-defender-defcon.pdf} \label{df_tt_d}
		}
		\subfloat[Defender's EE]{
			\includegraphics[width=.33\textwidth]{figures/fig17-eae-defender-defcon.pdf} \label{df_eae_d}
		}
		\caption{Evolution metrics measured for the defender (i.e., Snort) with the DEFCON dataset ($x$-axis unit: day).}
		\label{fig:DF-Latency-Defender}
	\end{figure*}
}

In summary, the defender's agility is comparable to the attacker's. In addition, a frequent evolution does not necessarily result in an effective evolution. This means that the defender's agility cannot be strictly related to defense effectiveness in overall. Therefore, we need to measure cyber agility separately from defense effectiveness.

\subsection{Case Study with DEFCON Dataset}

Fig.~\ref{fig:DF-Latency-Defender} demonstrates the measurement of evolution metrics when replaying the DEFCON dataset against Snort. Since we have three distinct DEFCON CTF attack traffic datasets (i.e., one capture per year), $\A_{t'_j}$, for $j=0, 1, 2$, where $t'_0=238$, $t'_1=609$, and $t'_2=973$, we can naturally treat each of them as an attack generation.

{\bf Analysis on Fig.~\ref{fig:DF-Latency-Defender}~(a):} This figure plots the sample of the defender's GT and the sample of the defender's EGT corresponding to the data available, where each release of Snort is treated as a generation. Since there are only three attack generations where each generation serves as a reference point for defining the corresponding metric $\EGT(\D,t'_j)$ for $j=0, 1$ and $2$, we have only three measurements of $\EGT(\D,t)$ for $t=238, 609, 973$, showing only three red bars. Notice that the blue bars are the same as in Fig.~\ref{fig:Honeypot-Latency}~(a). Defense evolution actions responding to $\A_{238}$ and $\A_{958}$ took relatively less generation time, showing 49 days and 69 days, respectively. However, the defense evolution took more than one year (i.e., 441 days) to respond to $\A_{609}$. This may be because of a small set of samples to draw conclusions about the evolution rates of these two parties.

\noindent{\bf Analysis on Fig.~\ref{fig:DF-Latency-Defender}~(b):} This figure exhibits a the defender's LBT. In Fig.~\ref{fig:DF-Latency-Defender}~(b), the purple curve indicates the average of $\LBT(\D)$ samples, which are shown in green, red, and blue. The security threshold, $\varepsilon$ (shown in teal), is set at a true-positive rate of 12\%; in general, $\varepsilon$ is the minimum acceptable value for the metric chosen (i.e. true-positive rate). The average $\LBT(\D)$ falls below $\varepsilon$ at $x = 1.14$ years, meaning that for the threshold chosen, the defender lags behind the attacker by 1.14 years.

\noindent{\bf Analysis on Fig.~\ref{fig:DF-Latency-Defender}~(c):} This figure addresses the sample of defender's EE based on the three attack generations with $t'=238, 609, 973$, resulting in three curves, each corresponding to one of the three $\A_{t'}$'s. We expect that for a fixed reference attack $\A_{t'}$, the defense effectiveness $\D_{t}(\A_{t'},M)$ should increase over time. However, this is not universally true because the Snort rules are not monotonically increasing (i.e., some rules are occasionally deleted in order to reduce false-positives or replace poorly written rules). This explains why the effectiveness is not monotonically increasing against each given attack generation.

{\bf Results Analysis:} Based on the result in Fig.~\ref{fig:DF-Latency-Defender}~(c), we examine the alert types from DEFCON 22. We learn that the attackers in DEFCON 22 primarily attacked protocols for which the Snort Preprocessor Plugins had more reliable effectiveness, such as HTTP Inspect pre-processors (relative to those in DEFCON 21). The subsequent year, DEFCON 23, shows the worst detection across all defensive generations. Because this attack generation was more effective than the corresponding defensive generations, we can conclude that the attacker outmaneuvers the defenders (i.e., the Snort community) in this case.

From the defender's EE with reference to DEFCON 22, we also observe that the EE for $A_{609}$ (DEFCON 22) increases sharply when defense $\D_{208}$ is released. In this case, the defender predicted the attack generation. One situation which could cause this is that the defender (i.e., the Snort community) saw a proof-of-concept exploit, and then evolved in order to mitigate this exploit. Later, the exploit became widespread in the wild, and was used by the attackers. However, the exploit was not successful when the attackers finally adopted it because the defender had already prepared for it. The exploit eventually fell out of popularity, so it ceased to show up in the next attack generation. This example would explain why the change appears in DEFCON 22 data, but not DEFCON 21 or DEFCON 23 data.

In summary, Snort exhibits a lower responsiveness to human-launched attacks, which are presumably most prevalent in DEFCON CTF competitions. However, the \emph{static effectiveness}, interpreted as a static metric sequence $\D_t(\A_t,M)$ for $t=0,1,\ldots,T$, is actually higher than its counterpart, the dataset observed by the honeypot, which is more likely to contain mostly automated attacks. This is partly because Snort has some aspects of the proactive defense capability. From these observations, \emph{cyber agility} and \emph{static effectiveness} need to be separately investigated.

\subsection{Discussion}
For the honeypot dataset, the mean of $\D_t(\A_t,tp)$ over the time horizon is $8.11\%$. For the DEFCON dataset, the mean of $\D_t(\A_t,tp)$ over $t\in \{238,609,973\}$ is $17.67\%$. This means that Snort is not effective overall. However, we suspect that the lower true-positive rate of Snort against the honeypot dataset is largely because these low-interaction honeypots limit the depth to which the attacker may penetrate. If the dataset were collected by a high-interaction honeypot, the true-positive rate would have been higher. Snort appears to be more effective in detecting reconnaissance activities (mainly shown in the honeypot dataset) than detecting exploitation activities (mainly shown in the DEFCON CTF activities). This is why Snort appears to evolve more efficiently against the attacks observed by the honeypot.


In summary, according to the perspective of the \emph{static effectiveness}, Snort is more effective in detecting attacks largely launched by human attackers than detecting attacks observed by honeypots. One caveat is that the attacks observed by low-interaction honeypots are not semantically rich enough because they capture only a limited interaction with attackers. Defense generations appear to be more effective in offsetting attacks observed by honeypots than offsetting attacks launched by human attackers. Nevertheless, Snort exhibits a potential proactive defense capability, as discussed earlier. Therefore, \emph{cyber agility} needs to be separately investigated from \emph{static effectiveness}.
Another application of these metrics is the following:
an attacker may be interested in predicting how long a zero-day attack will be usable before an effective defense will be deployed. For example, the attacker can calculate the expected EGT in the response to zero-day attacks. This also helps the defender to estimate the attacker's expectation of EGT, which may be leveraged to launch advanced defense (e.g., deception).

\section{Limitations} \label{sec: limitations}
\label{sec:limitation}
In this section, we discuss the limitations of our study.
Addressing these limitations will guide us to strengthen and refine the current metrics framework and other related metrics research.

First, the metrics require the defender to record the network traffic and/or computer execution traces in order to measure $\A_t(\D_{t'})$ in retrospect, where $t<t'$, $t=t'$ or $t>t'$. This may not always be feasible, especially for high speed networks that generate a large volume of network traffic or complex applications that may incur concurrent executions. Nevertheless, this appears to be the only way to measure the response to new or zero-day attacks.

Second, the used datasets are not ideal because they lack rich semantics (i.e., the low-interaction honeypot dataset) or continuity over a long period of time (i.e., the DEFCON dataset). Nevertheless, we are able to validate most metrics using one dataset or the other (with the exception of LBT). Because the datasets complement each other, these experiments sufficiently demonstrate the usefulness of the framework.

Third, the TT metric aims to correlate the effects of adversarial actions between the two parties. In the datasets used, this may not accurately represent the causality of their evolution. For example, in the DEFCON dataset, Snort was not updated during a competition, so evolution by the attacker may not have been caused by a change in Snort's rules. Nevertheless, an approximation to the ``ground-truth'' evolution generation may, as discussed in Section \ref{sec:discussion}, be practically useful enough.

Fourth, our study represents only the first step towards the ultimate goal of measuring cyber agility. Even if many limitations exist as mentioned above, our case study clearly shows that a detection tool, like Snort, evolves effectively and has proactive defense capability. These have not been studied in the literature.

Fifth, the proposed framework contains a set of metrics that may only capture some aspects of cyber attack and defense evolution. The framework may require further investigation to fully establish a sense of \emph{completeness}, which is important because any security metric of interest can be derived from a \emph{complete} set of metrics.

\section{Conclusion}
\label{sec:conclusion}
We have presented a suite of metrics to measure cyber agility by estimating the degree of attack and defense generational evolution. The proposed set of metrics includes generation time, effective generation time, triggering time (from detection to perform an adaptive defense), evolutionary effectiveness, lagging behind time, relative generational impact, and aggregated generational impact. These metrics mainly focus on measuring the timeliness and effectiveness in order to capture the core concepts of agility. 
We demonstrated the measurement of these metrics using the two real-world datasets (i.e., honeypots and DEFCON) which were tested using Snort. We discussed the underlying meanings of these metrics as well as their implications. 

The metrics proposed in this paper can provide valuable insights for defense strategization because they allow a defender to measure several aspects of cyber agility. These aspects include the defender's own responsiveness to attacks over time, the identification of which defense changes have been targeted by attackers, and the ongoing effectiveness of the defender against a single or set of attackers. Insights derived from these metrics will enable the defender to become more responsive, more targeted and more effective in their competition to outmaneuver attackers.
As one example, we mention that the ability to measure security effectiveness in retrospect makes it possible to characterize why the defense failed against new or zero-day attacks. This may lead to insights into how the failure may be prevented in the future. As another example, we mention that being able to tell the attackers apart would allow the defender to use our metrics to tell which attackers are more agile than others. This would suggest the defender to prioritize the defense correspondingly (e.g., paying special attention to the more agile attackers). As yet another example, we mention that the defender can use our metrics to tell which defense changes have been particularly targeted by attackers. This would suggest the defender to use more advanced defense techniques (e.g., deception) to offset the attacker's agility against these defense changes.

\smallskip

\noindent{\bf Acknowledgment.} We thank the reviewers for their insightful comments that guided us in improving the paper. For example, the term \emph{generation} was suggested to replace our original term of \emph{adaptation} because the former can more broadly accommodate attack and defense updates that are not necessarily incurred by a specific opponent move, effectively making the metric framework more widely applicable.

This research was supported in part by the US Department of Defense (DoD) through the office of the Assistant Secretary of Defense for Research and Engineering (ASD (R\&E)), ARO Grant \#W911NF-17-1-0566, ARL Grant \#W911NF-17-2-0127, and NSF Grant \#1814825. The views and opinions of the authors do not reflect those of the US DoD, ASD (R\&E), Air Force Research Laboratory, US Army, or NSF. Approved for Public Release; Distribution Unlimited: 88ABW-2019-1731 Dated 15 April 2019.

\bibliographystyle{ieeetr}
\bibliography{metrics}

\begin{IEEEbiographynophoto}{Jose David Mireles} is cybersecurity researcher and graduate of University of Texas at San Antonio (UTSA) where he received his M.S in Computer Science in 2017.  As a graduate student at UTSA, his research was focused on the extraction and attribution of attack signatures from large data sets, and also the measurement and analysis of attacker-defender interactions (cyber agility). He is a recipient of the National Science Foundation's Scholarship for Service program, and a RSA Conference Security Scholar (2017). 
\end{IEEEbiographynophoto}

\begin{IEEEbiographynophoto}{Eric Ficke} is a Cyber Security researcher at the University of Texas at San Antonio. He is currently pursuing his PhD in Computer Science. His existing research covers intrusion detection and cyber security metrics.
\end{IEEEbiographynophoto}

\begin{IEEEbiographynophoto}{Patrick Hurley} is a Principal Computer Engineer and Program Manager at the Air Force Research Laboratory in Rome, New York. He received a Master of Science degree in Computer Information Science from SUNY Institute of Technology. Mr. Hurley is currently the lead on a major AFRL program: Trusted and Resilient Systems. This program is focused on fighting through cyber-attacks while maintaining mission essential functions. For over 30 years, Mr. Hurley has been the lead technical agent for DARPA on
cyber defense programs that focus on survivable architectures, adaptive
security, and advanced distributed systems technologies that lead to more
agile and better managed systems. He is a member of the IEEE and has over 60
peer-reviewed technical papers in leading journals and conferences.
\end{IEEEbiographynophoto}

\begin{IEEEbiographynophoto}{Jin-Hee Cho} is currently an associate professor in the Department of Computer Science at Virginia Tech since Aug. 2018. Prior to joining the Virginia Tech, she worked as a computer scientist at the U.S. Army Research Laboratory (USARL), Adelphi, Maryland, since 2009. Dr. Cho has published over 110 peer-reviewed technical papers in leading journals and conferences in the areas of trust management, cybersecurity, metrics and measurements, network performance analysis, resource allocation, agent-based modeling, uncertainty reasoning and analysis, information fusion / credibility, and social network analysis. She received the best paper awards in IEEE TrustCom’2009, BRIMS’2013, IEEE GLOBECOM’2017, 2017 ARL’s publication award, and IEEE CogSima 2018. She is a winner of the 2015 IEEE Communications Society William R. Bennett Prize in the Field of Communications Networking. In 2016, Dr. Cho was selected for the 2013 Presidential Early Career Award for Scientists and Engineers (PECASE), which is the highest honor bestowed by the US government on outstanding scientists and engineers in the early stages of their independent research careers. Dr. Cho earned MS and PhD degrees in computer science from the Virginia Tech in 2004 and 2008, respectively. She is a senior member of the IEEE and a member of the ACM.
\end{IEEEbiographynophoto}

\begin{IEEEbiographynophoto}{Shouhuai Xu}
is a Full Professor in the Department of Computer Science, University of Texas at San Antonio. He is the Founding Director of the Laboratory for Cybersecurity Dynamics (\url{http://www.cs.utsa.edu/~shxu/LCD/index.html}). He coined the notion of Cybersecurity Dynamics as a candidate foundation for the emerging science of cybersecurity. His research interests include the three pillar thrusts of Cybersecurity Dynamics: first-principle cybersecurity modeling and analysis (the $x$-axis, to which the present paper belongs); cybersecurity data analytics (the $y$-axis); and cybersecurity metrics (the $z$-axis).  He co-initiated the International Conference on Science of Cyber Security (http://www.sci-cs.net/) and the ACM Scalable Trusted Computing Workshop (ACM STC). He is/was a Program Committee co-chair of SciSec'19, SciSec'18, ICICS'18, NSS'15 and Inscrypt'13. He is/was an Associate Editor of IEEE Transactions on Dependable and Secure Computing (IEEE TDSC), IEEE Transactions on Information Forensics and Security (IEEE T-IFS), and IEEE Transactions on Network Science and Engineering (IEEE TNSE). He received his PhD in Computer Science from Fudan University.
\end{IEEEbiographynophoto}

\end{document}